\documentclass[aps,prb,showpacs,twocolumn,superscriptaddress]{revtex4-1}%
\usepackage{graphics,bm}
\usepackage{amssymb}
\usepackage{epstopdf}
\usepackage[usenames]{color}
\usepackage{upgreek}
\usepackage{amsmath}
\usepackage{xcolor}
\usepackage{hyperref}
\usepackage{amsfonts}
\usepackage{graphicx}%
\providecommand{\U}[1]{\protect\rule{.1in}{.1in}}
\setcitestyle{numbers,square}
\newcommand\Tstrut{\rule{0pt}{2.6ex}}

\renewcommand{\vec}[1]{\bm{ {\bf #1}}}

\begin{document}

\title{Magnon spin transport driven by the magnon chemical potential in a magnetic insulator}
\author{L.J. Cornelissen}
\affiliation{Physics of Nanodevices, Zernike Institute for Advanced Materials, University
	of Groningen, Nijenborgh 4, 9747 AG Groningen, The Netherlands }
\author{K.J.H. Peters}
\affiliation{Institute for Theoretical Physics and Center for Extreme Matter and Emergent
	Phenomena, Utrecht University, Leuvenlaan 4, 3584 CE Utrecht, The Netherlands }

\author{G.E.W. Bauer}
\affiliation{Institute for Materials Research and WPI-AIMR, Tohoku University, Sendai, Japan}
\affiliation{Kavli Institute of NanoScience, Delft University of Technology, Delft, The
	Netherlands }
\author{R.A. Duine}
\affiliation{Institute for Theoretical Physics and Center for Extreme Matter and Emergent
	Phenomena, Utrecht University, Leuvenlaan 4, 3584 CE Utrecht, The Netherlands }
\affiliation{Department of Applied Physics, Eindhoven University of Technology, PO Box 513, 5600 MB Eindhoven, The Netherlands}
\author{B.J. van Wees}
\affiliation{Physics of Nanodevices, Zernike Institute for Advanced Materials, University
	of Groningen, Nijenborgh 4, 9747 AG Groningen, The Netherlands }

\begin{abstract}
We develop a linear-response transport theory of diffusive spin and heat
transport by magnons in magnetic insulators with metallic contacts. The
magnons are described by a position dependent temperature and chemical
potential that are governed by diffusion equations with characteristic
relaxation lengths. Proceeding from a linearized Boltzmann equation, we derive
expressions for length scales and transport coefficients. For yttrium iron
garnet (YIG) at room temperature we find that long-range transport is
dominated by the magnon chemical potential. We compare the model's results with
recent experiments on YIG with Pt contacts [L.J. Cornelissen, \emph{et al.},
Nat. Phys. \textbf{11}, 1022 (2015)] and extract a magnon spin
conductivity of $\sigma_{m}=5\times10^{5}$ S/m. Our results for the spin
Seebeck coefficient in YIG agree with published experiments. We conclude that
the magnon chemical potential is an essential ingredient for energy and spin
transport in magnetic insulators.

\end{abstract}

\maketitle

\section{Introduction}

\label{sec:introduction} The physics of diffusive magnon transport in magnetic
insulators, first investigated by Sanders and Walton\ \cite{Sanders1977}, has
been a major topic in spin caloritronics since the discovery of the spin
Seebeck effect (SSE) in YIG%
$\vert$%
Pt bilayers \cite{Uchida2010,Bauer2012,Adachi2013}. This transverse voltage
generated in platinum contacts to insulating ferromagnets under a temperature
gradient can be explained by thermal spin pumping caused by a temperature
difference between magnons in the ferromagnet and electrons in the platinum
\cite{Xiao2010,Adachi2013,PhysRevB.88.094410,PhysRevB.89.014416}. The magnons
and phonons in the bulk ferromagnet are considered as two weakly interacting
subsystems, each with their own temperature \cite{Sanders1977}. Hoffman
\emph{et al.} explained the spin Seebeck effect in terms of the stochastic
Landau-Lifshitz-Gilbert equation with a noise term that follows the phonon
temperature \cite{PhysRevB.88.064408}.

Recently, diffusive magnon spin transport over large distances has been
observed in YIG that was driven either electrically \cite{Cornelissen2015,
Goennenwein2015}, thermally \cite{Cornelissen2015} or optically
\cite{Giles2015}. Notably, our observation of electrically driven magnon spin transport was recently confirmed in a Pt\textbar{}YIG\textbar{}Pt trilayer geometry\cite{Wu2016,Li2016}. Here we argue that previous theories cannot explain these
observations, and therefore do not capture the complete physics of magnon
transport in magnetic insulators. We present arguments in favor of a
non-equilibrium magnon chemical potential and work out the consequences for
the interpretation of experiments.

Magnons are the elementary excitations of the magnetic order parameter. Their
quantum mechanical creation and annihilation operators fulfill the boson
commutation relations as long as their number is sufficiently small. Just like
photons and phonons, magnons at thermal equilibrium are distributed over
energy levels according to Planck's quantum statistics for a given temperature
$T$. This is a Bose-Einstein distribution with zero chemical potential,
because the energy and therefore magnon number is not
conserved. Nevertheless, it is well established that a magnon chemical
potential can parametrize a long-living \textit{non-equilibrium} magnon state.
For instance, parametric excitation of a ferromagnet by microwaves generates
high energy magnons that thermalize much faster by magnon-conserving exchange
interactions than that their number decays \cite{Serga2010}. The resulting
distribution is very different from a zero-chemical potential quantum or
classical distribution function, but is close to an equilibrium distribution
with a certain temperature and nonzero chemical potential. The breakdown of
even such a description is then indicative of the creation of a Bose (or, in
the case of pumping at energies much smaller than the thermal one,
Rayleigh-Jeans \cite{Ruckriegel2015}) condensate. This new state of matter has
indeed been observed \cite{Demokritov2006}. Here we argue that a magnon
chemical potential governs spin and heat transport not only under strong
parametric pumping, but also in the linear response to weak electric or
thermal actuation \cite{2015arXiv150501329D}.

The elementary magnetic electron-hole excitations of normal metals or spin
accumulation has been a very fruitful concept in spintronics \cite{Fabian2004}%
. Since electron thermalization is faster than spin-flip decay, a spin
polarized non-equilibrium state can be described in terms of two Fermi-Dirac
distribution functions with different chemical potentials and temperatures for
the majority and minority spins. We may distinguish the \emph{spin (particle)
accumulation} as the difference between chemical potentials from the
\emph{spin heat accumulation} as the difference between the spin temperatures
\cite{Dejene2013}. Both are vectors that are generated by spin injection and
governed by diffusion equations with characteristic decay times and lengths.
The spin heat accumulation decays faster than the spin particle accumulation,
since both are dissipated by spin-flip scattering, while the latter is inert
to energy exchanging electron-electron interactions. Here we proceed from the
premise that non-equilibrium states of the magnetic order can be described by
a Bose-Einstein distribution function for magnons that is parametrized by both
temperature and chemical potential, where the latter implies magnon number
conservation. We therefore define a \emph{magnon heat accumulation} $\delta
T_{m}$ as the difference between the temperature of the magnons and that of
the lattice. The chemical potential $\mu_{m}$ then represents the \emph{magnon
spin accumulation}, noting that this definition differs from that by Zhang and
Zhang \cite{PhysRevLett.109.096603}, who define a magnon spin accumulation in
terms of the magnon density. The crucial parameters are then the relaxation
times governing the equilibration of $\delta T_{m}$ and $\mu_{m}$. When the
magnon heat accumulation decays faster than the magnon particle accumulation,
previous theories for magnonic heat and spin transport should be doubted
\cite{Sanders1977,Xiao2010,Flipse2014,PhysRevB.88.094410,PhysRevB.89.014416}.
The relaxation times are governed by the collision integrals that include
inelastic (one, two and three magnon scatterings involving phonons) and
elastic two and four-magnon scattering processes. At room temperature, two-magnon scattering due to disorder is likely to be negligibly small
compared to phonon scattering. Four-magnon scattering only redistributes the
magnon energies, but does not lead to momentum or energy loss of the magnon
system. Processes that do not conserve the number of magnons are caused by
either dipole-dipole or spin-orbit interaction with the lattice and should be less important than the
magnon-conserving ones for
high quality magnetic materials such as YIG. At room temperature, the magnon spin accumulation is
then essential to describe diffusive spin transport in ferromagnets.

Here we revisit the linear response transport theory for magnon spin and heat
transport, deriving the spin and heat currents in the bulk of the magnetic
insulator as well as across the interface with a normal metal contact. The
magnon transport is assumed to be diffusive. Formally we are then limited to the
regime in which the thermal magnon wavelength $\Lambda$ and the magnon mean
free path $\ell$ (the path length over which magnon momentum is conserved) are
smaller than the system size $L$. The wavelength of magnons in YIG in a
simple parabolic band model and is a few nanometers at room temperature.{
Boona \emph{et al.} \cite{PhysRevB.90.064421} find that $\ell$ at room
temperature is of the same order. As in electron transport in magnetic
multilayers, scattering at rough interfaces is likely to render a diffusive
picture valid even when the formal conditions for diffusive bulk transport are
not met. Under the assumptions that magnons thermalize efficiently and that
the mean-free path is dominated by magnon-conserving scattering by phonons or
structural and magnetic disorder, we find that the magnon chemical potential
is required to harmonize theory and experiments on magnon spin transport
\cite{Cornelissen2015}. }

{This paper is organized as follows: We start with a brief review of diffusive
charge, spin and heat transport in metals in
Sec.~\ref{subsec:spintransportnormalmetals}. In
Sec.~\ref{subsec:spintransportmagneticinsultors}, we derive the linear
response expressions for magnon spin and heat currents, starting from the
Boltzmann equation for the magnon distribution function. We proceed with
boundary conditions at the Pt%
$\vert$%
YIG interface in Sec.~\ref{subsec:interfacialspincurrents}. In
Sec.~\ref{subsec:lengthscales} we provide estimates for relaxation lengths and
transport coefficients for YIG. The transport equations are analytically
solved for a one-dimensional model (longitudinal configuration) in
Sec.~\ref{subsec:onedimension}. In Sec.~\ref{subsec:twodimensions} we
implement a numerical finite-element model of the experimental geometry and we
compare results with experiments in Sec.~\ref{subsec:comparison}. We apply our
model also to the (longitudinal) spin Seebeck effect in Sec.~\ref{subsec:sse}.
A summary and conclusions are given in Sec.~\ref{sec:conclusions}.
\begin{figure}[h]
\includegraphics[width=8.5cm]{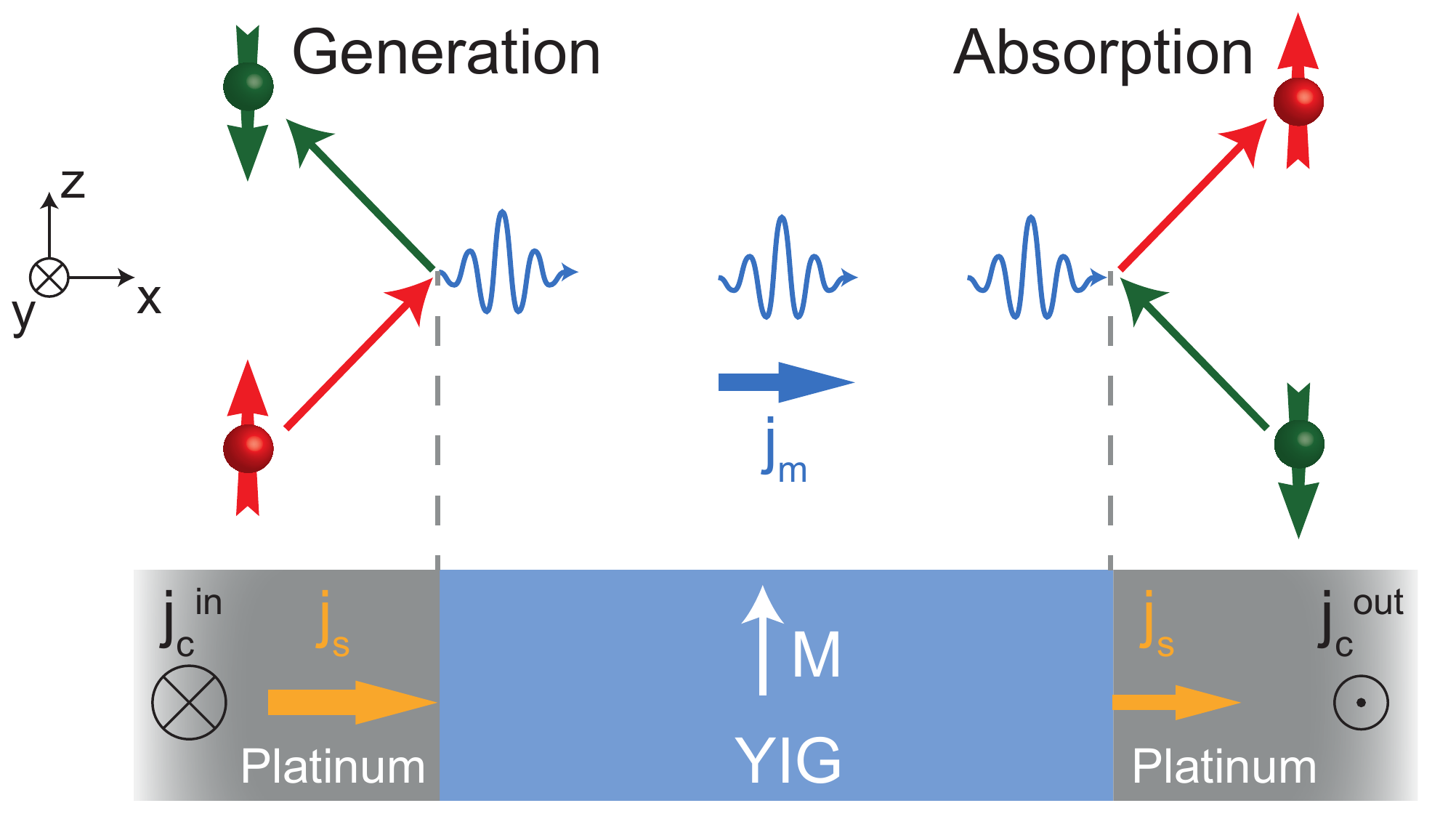}\caption{Schematic of the 1D geometry \cite{PhysRevLett.109.096603, Li2016}. A charge current $j_{c}^{in}$ is sent
through the left platinum strip along $+\vec{y}$. This generates a spin
current $j_{s}=j_{\mathrm{xz}}=\theta j_{c}^{in}$ towards the YIG$|${}Pt
interface and a spin accumulation, injecting magnons into the YIG with spin
polarization parallel to the magnetization $\vec{M}$. The magnons diffuse
towards the right YIG$|${}Pt interface, where they excite a spin accumulation
and spin current into the contact. Due to the inverse spin Hall effect, this
generates a charge current $j_{c}^{out}$ along the $-\vec{y}$ direction. Note
that if $\vec{M}$ is aligned along $-\vec{z}$, magnons are absorbed at the
injector and created at the detector.}%
\label{fig:geometry1D}%
\end{figure}}

\section{Theory}

\label{sec:model} We first review the diffusion theory for electrical magnon
spin injection and detection as published by one of us in
\cite{2015arXiv150501329D,2015arXiv151005316F}. By introducing the magnon
chemical potential this approach can disentangle spin and heat transport in
contrast to earlier treatments based on the magnon density
\cite{PhysRevLett.109.096603} or magnon temperature
\cite{Sanders1977,PhysRevB.88.094410,PhysRevB.89.014416, Xiao2010} only. We
initially focus on the one-dimensional geometry in Fig.~\ref{fig:geometry1D}
with two normal metal (Pt) contacts to the magnetic insulator YIG. We express
the spin currents in the bulk of the normal metal contacts and magnetic
spacer, and the interface. While Ref. \cite{2015arXiv150501329D} focussed on
the chemical potential, here we include the magnon temperature as well. At low
temperatures the phonon specific heat has been reported to be an order of
magnitude larger than the magnon one \cite{PhysRevB.90.064421}. The
room-temperature phonon mean free path (that provides an upper bound for the
phonon collision time) of a few nm \cite{PhysRevB.90.064421} corresponds to a
sub-picosecond transport relaxation time for sound velocities of
$10^{3}-10^{4}$ m/s. From the outset, we therefore take the phonon heat
capacity to be so large and the phonon mean free path and collision times so
short that the phonon distribution is not significantly affected by the
magnons. The phonon temperature $T_{p}$ is assumed to be either a fixed
constant or, in the spin Seebeck case, to have a constant gradient. For
simplicity, we also disregard the finite thermal (Kapitza) interface heat
resistance of the phonons \cite{Cahill2003}.

\subsection{Spin and heat transport in normal metals}

\label{subsec:spintransportnormalmetals} There is much evidence that spin
transport in metals is well described by a spin diffusion approximation.
Spin-flip diffusion lengths of the order of nanometers reported in platinum
betray the existence of large interface contributions \cite{Wang2015}, but the
parameterized theory describes transport well \cite{Chen2015}. The charge
($j_{c,\alpha}$), spin ($j_{\alpha\beta}$) and heat ($j_{Q,\alpha}$) current
densities in the normal metals, where the spin polarization is defined in the
coordinate system of Fig.~\ref{fig:geometry1D}, are given by (see e.g.
\cite{PhysRevB.90.014428}) \begin{widetext}
\begin{eqnarray}
\label{eq:spindrifteqsmetal}
j_{c,\alpha} &=& \sigma_e \partial_\alpha\mu_e -
\sigma_e S \partial_\alpha T_e - \frac{\sigma_{\rm SH}}{2} \epsilon_{\alpha
\beta\gamma}\partial_\beta\mu_\gamma~, \nonumber\\
\frac{2e}{\hbar} j_{\alpha\beta}  &=& -\frac{\sigma_e}{2}  \partial_{\alpha}\mu_{\beta}- \sigma_{\rm SH} \epsilon_{\alpha\beta\gamma}  \partial_{\gamma}\mu_e  -\sigma_{\rm SH} S_{\rm SN}
\epsilon_{\alpha\beta\gamma} \partial_\gamma T_e~, \nonumber\\
j_{Q,\alpha} &=& -
\kappa_e \partial_\alpha T_e - \sigma_e P \partial_\alpha\mu_e - \frac
{\sigma_{\rm SH}}{2}  P_{\rm SN} \epsilon_{\alpha\beta\gamma} \partial_\beta
\mu_\gamma~.
\end{eqnarray}
\end{widetext} Here, $\mu_{e}$, $T_{e}$, and $\mu_{\alpha}$ denote the
electrochemical potential, electron temperature, and spin accumulation,
respectively. The subscripts $\alpha,\beta,\gamma\in\{x,y,z\}$ are Cartesian
components in the coordinate system in Fig. \ref{fig:geometry1D}, $\alpha$
indicating current direction and $\beta$ spin polarization. $\epsilon
_{\alpha\beta\gamma}$ is the Levi-Civita tensor and the summation convention
is assumed throughout. The charge, spin, and heat current densities are
measured in units of A/m$^{2}$, J/m$^{2}$ and W/m$^{2},$ respectively, while
both the electrochemical potential and the spin accumulation are in volts. The
charge and spin Hall conductivities are $\sigma_{e}$ and $\sigma_{\mathrm{SH}%
}$, both in units of S/m. Thermoelectric effects in metals are governed by the
Seebeck coefficient $S$ and Peltier coefficient $P=ST_{e}$. Similarly, we
allow for a spin Nernst effect via the coefficient $S_{\mathrm{SN}}$ and the
reciprocal spin Ettingshausen effect governed by $P_{\mathrm{SN}%
}=S_{\mathrm{SN}}T_{e}$. We assume, however, that spin-orbit coupling is weak
enough so that we can ignore spin swapping terms, i.e., terms of the form
$j_{\alpha\beta}\sim\partial_{\beta}\mu_{\alpha}$ and their Onsager reciprocal
\cite{PhysRevLett.103.186601}. The spin heat accumulation in the normal metal
and therefore spin polarization of the heat current are disregarded for
simplicity \cite{Dejene2013}. $\hbar$ and $-e$ are Planck's constant and the
electron charge. The continuity equation $\partial_{t}\rho_{e}%
+\boldsymbol{\nabla}\cdot\mathbf{j}_{e}=0$ expresses conservation of the
electric charge density $\rho_{e}$. The electron spin $\boldsymbol{\mu}$ and
heat $Q_{e}$ accumulations relax to the lattice at rates $\Gamma_{s\mu}$ and
$\Gamma_{QT},$ respectively:
\begin{align}
\partial_{t}s_{\beta}+\frac{1}{\hbar}\partial_{\alpha}j_{\alpha\beta}  &
=-2\Gamma_{s\mu}e\mu_{\beta}\nu~,\\
\partial_{t}Q_{e}+\boldsymbol{\nabla}\cdot j_{Q}  &  =-\Gamma_{QT}C_{e}\left(
T_{e}-T_{p}\right)  ~,
\end{align}
where the non-equilibrium spin density $s_{\beta}=2e\mu_{\beta}\nu$, $C_{e}$
is the electron heat capacity per unit volume, and $\nu$ the density of states
at the Fermi level. Inserting Eq.~(\ref{eq:spindrifteqsmetal}) leads to the
length scales $\ell_{s}=\sqrt{\sigma_{e}/\left(4e^{2}\Gamma_{s\mu}\nu\right)}$ and
$\ell_{\mathrm{ep}}=\sqrt{\kappa_{e}/\left(\Gamma_{QT}C_e\right)}$ governing the decay of the
electron spin and heat accumulations, respectively. At room temperature, these
are typically $\ell_{s}^{\mathrm{Pt}}=1.5$~nm, $\ell_{\mathrm{ep}%
}^{\mathrm{Pt}}=4.5$~nm for platinum \cite{Weiler2013,Flipse2014}, and
$\ell_{s}^{\mathrm{Au}}=35$~nm, $\ell_{\mathrm{ep}}^{\mathrm{Au}}=80$~nm for
gold \cite{Flipse2014,Isasa2015}.

\subsection{Spin and heat transport in magnetic insulators}

\label{subsec:spintransportmagneticinsultors} Magnonics traditionally focusses
on the low energy, long wavelength regime of coherent wave dynamics. In
contrast, the basic and yet not well tested assumption underlying the present
theory is diffusive magnon transport, which we believe to be appropriate for
elevated temperatures in which short-wavelength magnons dominate. Diffusion
should be prevalent when the system size is larger than the magnon mean free
path and magnon thermal wavelength (called magnon coherence length in
\cite{Xiao2010}). Magnons carry angular momentum parallel to the magnetization
($z$-axis). Oscillating transverse components of the angular momentum can be
safely neglected for system sizes larger than the magnetic exchange length, which is on the order of ten nanometer in YIG at low external magnetic fields \cite{PhysRevB.88.064408}.

Not much is known about the scattering mean-free path, but extrapolating the
results from Ref.~\cite{PhysRevB.90.064421} to room temperature leads to an
estimate of a few nm. Dipolar interactions affect mainly the long wavelength
coherent magnons that do not contribute significantly at room temperature.
Thermal magnons interact by strong and number-conserving exchange
interactions. In the Appendix the magnon-magnon scattering rate is estimated
as $(T/T_{c})^{3}k_{B}T/\hbar$ \cite{PhysRev.102.1217,PhysRevB.90.094409} or a
scattering time of $0.1$ ps for YIG with Curie temperature $T_{c}\sim500$ K at
room temperature $T=300$ K, where $T\approx T_{m}\approx T_{p}$. According to
the Landau-Lifshitz-Gilbert phenomenology \cite{1353448} the magnon decay rate
is $\alpha_{G}k_{B}T/\hbar$ \cite{PhysRevB.90.094409}, with Gilbert damping
constant $\alpha_{G}\approx10^{-4}\ll1$ for YIG. Hence, the ratio between the
scattering rates for magnon non-conserving to conserving processes is
$\alpha_{G}(T_{c}/T)^{3}\ll1$ at room temperature. These numbers justify the
second crucial premise of the present formalism, viz. very efficient, local
equilibration of the magnon system. Since a spin accumulation in general
injects angular momentum and heat at different rates, we need at least two
parameters for the magnon distribution $f$, i.e. an effective temperature
$T_{m}$ and a non-zero chemical potential (or magnon spin accumulation)
$\mu_{m}$ in the Bose-Einstein distribution function $n_{B}$
\begin{equation}
f(\mathbf{x},\epsilon)=n_{B}\left(  \mathbf{x},\epsilon\right)  =\left(
e^{\frac{\epsilon-\mu_{m}(\mathbf{x})}{k_{B}T_{m}(\mathbf{x})}}-1\right)
^{-1},
\end{equation}
where $k_{B}$ is Boltzmann's constant. Both magnon accumulations $T_{m}-T_{p}$
and $\mu_{m}$ vanish on in principle different length scales during diffusion.
Assuming an isotropic (cubic) medium, the magnon spin current ($\mathbf{j}%
_{m}$, in J/m$^{2}$) and heat current densities ($\mathbf{j}_{\mathrm{Q,m}}$,
in W/m$^{2}$) in linear response read%
\begin{equation}%
\begin{pmatrix}
\frac{2e}{\hbar}\mathbf{j}_{m}\\
\mathbf{j}_{\mathrm{Q,m}}%
\end{pmatrix}
=-%
\begin{pmatrix}
\sigma_{m} & L/T\\
\hbar L/2e & \kappa_{m}%
\end{pmatrix}%
\begin{pmatrix}
\boldsymbol{\nabla}\mu_{m}\\
\boldsymbol{\nabla}T_{m}%
\end{pmatrix}
,\label{eq:linearresponsespinheatmagnet}%
\end{equation}
where $\mu_{m}$ is measured in volts, $\sigma_{m}$ is the magnon spin
conductivity (in units of S/m), $L$ is the (bulk) spin Seebeck coefficient in
units of A/m, and $\kappa_{m}$ is the magnonic heat conductivity in units of
Wm$^{-1}$K$^{-1}$. Magnon-phonon drag contributions $\mathbf{j}_{m}%
,\mathbf{j}_{\mathrm{Q,}m}\propto\boldsymbol{\nabla}T_{p}$ are assumed to be
absorbed in the transport coefficients since $T_{m}\approx T_{p}$. The spin
and heat continuity equations for magnon transport read
\begin{equation}%
\begin{pmatrix}
\frac{\partial\rho_{m}}{\partial t}+\frac{1}{\hbar}\boldsymbol{\nabla}%
\cdot\mathbf{j}_{m}\\
\frac{\partial Q_{m}}{\partial t}+\boldsymbol{\nabla}\cdot\mathbf{j}%
_{\mathrm{Q,}m}%
\end{pmatrix}
=-%
\begin{pmatrix}
\Gamma_{\rho\mu} & \Gamma_{\rho T}\\
\Gamma_{Q\mu} & \Gamma_{Q T}
\end{pmatrix}%
\begin{pmatrix}
\mu_{m}\frac{\partial\rho_{m}}{\partial\mu_{m}}\\
C_{m}\left(  T_{m}-T_{p}\right)
\end{pmatrix}
,\label{eq:diffeqsmagnet}%
\end{equation}
in which $\rho_{m}$ is the non-equilibrium magnon spin density and $Q_{m}$ the
magnonic heat accumulation. $C_{m}$ is the magnon heat capacity per unit
volume. The rates $\Gamma_{\rho\mu}$ and $\Gamma_{QT}$ describe relaxation of
magnon spin and temperature, respectively. The cross terms (decay or
generation of spins by cooling or heating of the magnons and vice versa) are
governed by the coefficients $\Gamma_{\rho T}$ and $\Gamma_{Q\mu}$.
\begin{figure}[h]
\includegraphics[width=8.5cm]{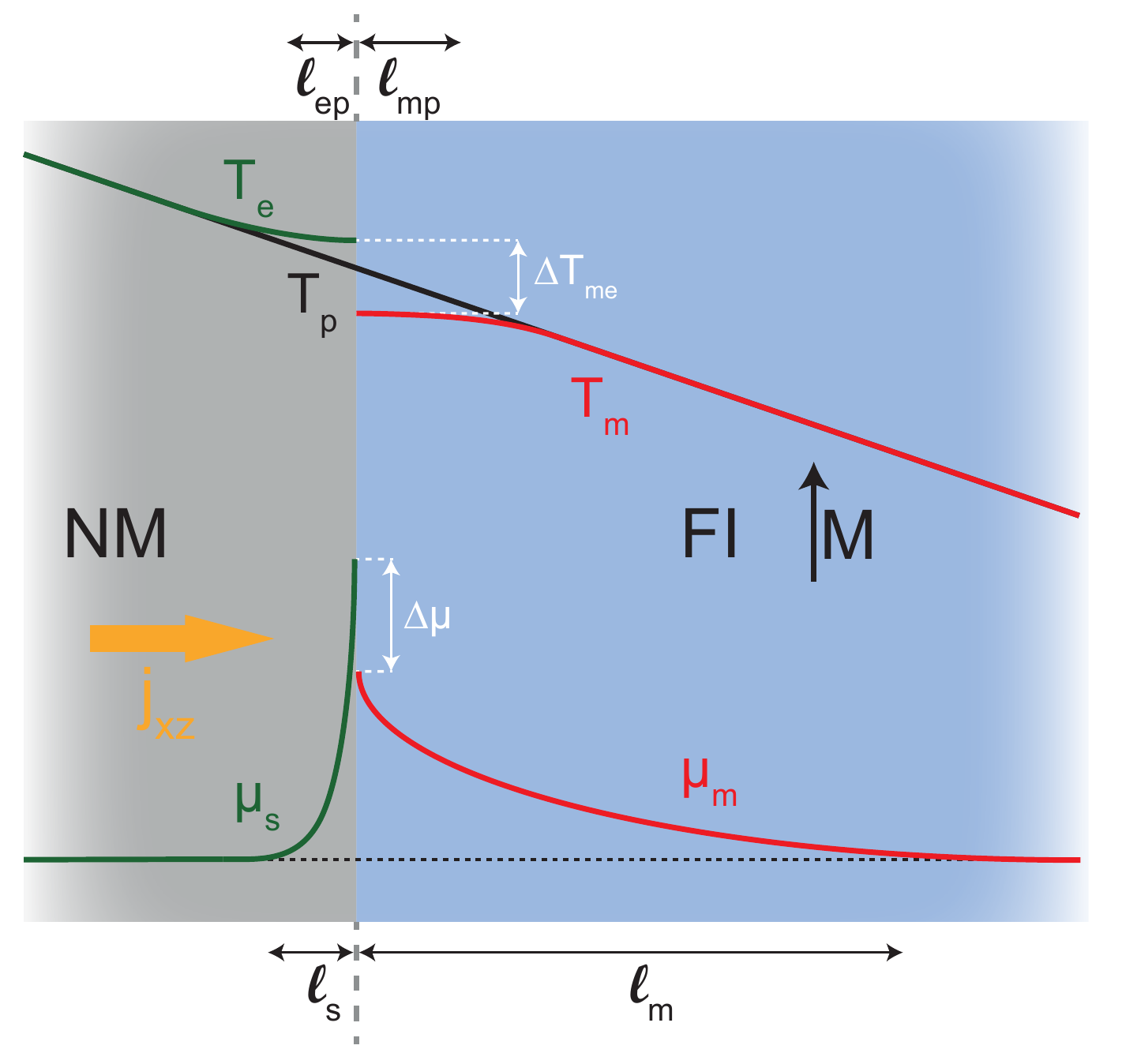}\caption{(Color
online) Length scales at normal metal$|${}ferromagnetic insulator (NM$|${}FI)
interfaces in Fig.~\ref{fig:geometry1D}. Assuming a constant gradient of the
phonon temperature $T_{p}$ and disregarding Joule heating, the electron
temperature $T_{e}$ and magnon temperature $T_{m}$ relax on length scales
$\ell_{\mathrm{ep}}$ and $\ell_{\mathrm{mp}}$. A significant phonon heat
(Kapitza) resistance would cause a step in $T_{p}$ at the interface. The spin
Hall effect in the normal metal drives a spin current $j_{\mathrm{xz}}$
towards the interface, which will be partially transmitted to the magnon
system (causing a non-zero magnon chemical potential in the FI) and partially
reflected back into the NM (causing a non-zero electron spin accumulation in
the NM). The electron spin accumulation $\mu_{s}=\mu_{z}$ and the magnon
chemical potential $\mu_{m}$ relax on length scales $\ell_{s}$ and $\ell_{m}$,
respectively.}%
\label{fig:relaxationlengths}%
\end{figure}Eqs.~(\ref{eq:linearresponsespinheatmagnet}%
)~and~(\ref{eq:diffeqsmagnet}) lead to the diffusion equations
\begin{gather}%
\begin{pmatrix}
e & \alpha_{\mu}k_{B}\\
e\alpha_{T}/k_{B} & 1
\end{pmatrix}%
\begin{pmatrix}
\nabla^{2}\mu_{m}\\
\nabla^{2}T_{m}%
\end{pmatrix}
=\nonumber\\%
\begin{pmatrix}
e/\ell_{m}^{2} & k_{B}/\left(  \ell_{\rho T}T^{2}\right)  \\
e/\left(  k_{B}\ell_{Q\mu}\mu_{m}^{2}\right)   & 1/\ell_{\mathrm{mp}}^{2}%
\end{pmatrix}%
\begin{pmatrix}
\mu_{m}\\
T_{m}-T_{p}%
\end{pmatrix}
,\label{eq:fullsteadystatemagnontransport}%
\end{gather}
with four length scales and two dimensionless ratios. $\ell_{m}=\sqrt
{\sigma_{m}/\left(  2e\Gamma_{\rho\mu}\right) \left(\frac{\partial \rho_m}{\partial \mu_m}\right)^{-1} }$ and $\ell
_{\mathrm{mp}}=\sqrt{\kappa_{m}/\left(\Gamma_{QT} C_m\right)}$ are the relaxation lengths of,
respectively, magnon chemical potential and temperature with equilibrium
values $\mu_{m}=0$ and $T_{m}=T_{p}$ (see Fig.~\ref{fig:relaxationlengths}).
The length scales $\ell_{\rho T}=\sqrt{k_{B}\sigma_{m}/\left(  2e^{2}\Gamma_{\rho
T} C_m \right)  }$ and $\ell_{Q\mu}=\sqrt{e \kappa_{m}/\left( \hbar k_{B}\Gamma_{Q\mu
}\right)  \left(\frac{\partial \rho_m}{\partial \mu_m} \right)^{-1} }$ arise from the non-diagonal cross terms. The dimensionless
ratio $\alpha_{\mu}=eL/\left(  k_{B}\sigma_{m}T_{p}\right)  $ is a measure for
the relative ability of chemical-potential and temperature gradients to drive
spin currents. Similarly, $\alpha_{T}=\hbar k_{B}L/\left(  2e\kappa
_{m}\right)  $ characterizes the magnon heat current driven by chemical
potential gradients relative to that driven by temperature gradients.

\subsection{Interfacial spin and heat currents}

\label{subsec:interfacialspincurrents} The electron and magnon diffusion
equations are linked by interface boundary conditions. Spin currents and
accumulations are parallel to magnetization direction of the ferromagnet along
the $z$-direction. We assume that the exchange coupling dominates the coupling
between electrons and magnons across the interface. A perturbative treatment
of the exchange coupling at the interface leads to the spin
current~\cite{bender2012,PhysRevB.91.140402}
\begin{align}
j_{s}^{\mathrm{int}}  &  =-\frac{\hbar\,g^{\uparrow\downarrow}}{2e^{2}\pi
s}\int d\epsilon D(\epsilon)\left(  \epsilon-e\mu_{z}\right)
\nonumber\label{eq:spincurrentinterface}\\
&  \times\left[  n_{B}\left(  \frac{\epsilon-e\mu_{m}}{k_{B}T_{m}}\right)
-n_{B}\left(  \frac{\epsilon-e\mu_{z}}{k_{B}T_{e}}\right)  \right]  ~,
\end{align}
where $g^{\uparrow\downarrow}$ is the real part of the spin mixing conductance
in S/m$^{2}$, $s=\mathrm{S}/\mathrm{a}^{3}$ the equilibrium spin density of the magnetic insulator and $\mathrm{S}$ is the total spin in a unit
cell with volume $\mathrm{a}^{3}$. The density of states of magnons
$D(\epsilon)=\sqrt{\epsilon-\Delta}/\left(  4\pi^{2}J_{s}^{3/2}\right)  $ for
a dispersion $\hbar\omega_{\mathbf{k}}=J_{s}\mathbf{k}^{2}+\Delta$. The spin
wave gap $\Delta$ is governed by the magnetic anisotropy and the applied
magnetic field. In soft ferromagnets such as YIG $\Delta\sim1$ K, which we
disregard in the following since we focus on effects at room temperature (see
e.g. Ref.~\cite{PhysRevB.88.064408}). The heat current is given by inserting
$\epsilon/\hbar$ into the integrand of Eq.~(\ref{eq:spincurrentinterface}).

Linearizing the above equation we find the spin and heat currents across the
interface \cite{2015arXiv150501329D}

\begin{widetext}
\begin{equation}
\label{eq:interfacecurrents}
\begin{pmatrix}
j_s^{\rm int} \\
j_Q^{\rm int}
\end{pmatrix} =
\frac{3\hbar\,g^{\uparrow\downarrow}}{4e^2 \pi s \Lambda^3}
\begin{pmatrix}
e\,\zeta\left(3/2\right) & \frac{5}{2}k_B\zeta\left(5/2\right) \\
\frac{5}{2}\frac{ek_B T}{\hbar}\zeta\left(5/2\right) & \frac{35}{4}\frac{k_B^2 T}{\hbar}\zeta\left(7/2\right)
\end{pmatrix}
\begin{pmatrix}
\mu_z-\mu_m \\
T_e - T_m
\end{pmatrix}.
\end{equation}
\end{widetext} $\Lambda=\sqrt{4\pi J_{s}/\left(  k_{B}T\right)  }$ is the
magnon thermal (de Broglie) wavelength (the factor $4\pi$ is included for
convenience). These expressions agree with those derived from a stochastic
model \cite{Xiao2010} after correcting numerical factors of the order of
unity. In YIG at room temperature $\Lambda\sim1\,$nm. The term proportional to
$\mu_{z}$ corresponds to the spin transfer (absorption of spin current by the
fluctuating magnet), while that proportional to $\mu_{m}$ is the spin pumping
contribution (emission of spin current by the magnet). The prefactor
$\sim1/\left(  s\Lambda^{3}\right)  $ can be understood by noting that
$s\Lambda^{3}$ is the effective number of spins in the magnetic insulator that
has to be agitated and appears in the denominator of
Eq.~(\ref{eq:interfacecurrents}) as a mass term. In the macrospin
approximation this term would be replaced by the total number of spins in the magnet.

From Eq.~(\ref{eq:interfacecurrents}) we identify the effective spin mixing
conductance $g_{s}$ that governs the transfer of spin across the interface by
the chemical potential difference $\Delta\mu=\mu_{z}-\mu_{m}$. In units of
S/m$^{2}$
\begin{equation}
g_{s}=\frac{3\,\zeta\left(  \frac{3}{2}\right)  }{2\pi s}\frac{g^{\uparrow
\downarrow}}{\Lambda^{3}}. \label{eqn:gs}%
\end{equation}
Using the material parameters for YIG from Tab.~\ref{tab:modelparameters} and the expression for the
thermal De Broglie wavelength given above, we find $g_{s}=0.06g^{\uparrow
\downarrow}$ at room temperature \cite{Xiao2015,Flipse2014}. $g_{s}$ scales
with temperature like $\sim(T/T_{c})^{3/2}$, but it should be kept in mind
that the theory is not valid in the limits $T\rightarrow T_{c}$ and
$T\rightarrow0.$ It is nevertheless consistent with the recently reported
strong suppression of $g_{s}$ at low temperatures \cite{Goennenwein2015}.

\subsection{Parameters and length scales}

\label{subsec:lengthscales}

In this section we present expressions for the transport parameters derived
from the linearized Boltzmann equation for the magnon distribution function
and present numerical estimates based on experimental data.

\subsubsection{Boltzmann transport theory}

\label{subsubsec:boltzmann}

Magnon transport as formulated in the previous section is governed by the
transport coefficients $\sigma_{m}$, $L$, $\kappa_{m}$, four length scales
$\ell_{m}$, $\ell_{\mathrm{mp}}$, $\ell_{\rho T}$ and $\ell_{Q\mu}$, and two
dimensionless numbers $\alpha_{\mu}$ and $\alpha_{T}$. In the Appendix we
derive these parameters using the linearized Boltzmann equation in the
relaxation time approximation. We consider four interaction events: i) elastic
magnon scattering by bulk impurities or interface disorder, ii) magnon
dissipation by magnon-phonon interactions that annihilate or create spin waves
and/or inelastic scattering of magnons by magnetic disorder, iii)
magnon-phonon interactions that conserve the number of magnons, and iv)
magnon-magnon scattering by magnon-conserving exchange scattering processes,
see also Sec.~{\ref{subsec:spintransportmagneticinsultors}}

The magnon energy and momentum dependent scattering times for these process
are $\tau_{\mathrm{el}}$, $\tau_{\mathrm{mr}}$, $\tau_{\mathrm{mp}}$, and
$\tau_{\mathrm{mm}}.$ At elevated temperatures they should be computed at
magnon energy $k_{B}T$ and momentum $\hbar/\Lambda$. Magnon-magnon
interactions that conserve momentum do not directly affect transport currents,
so the total relaxation rate is $1/\tau=1/\tau_{\mathrm{el}}+1/\tau
_{\mathrm{mr}}+1/\tau_{\mathrm{mp}}$.

The transport coefficients and length scales derived in the appendix are
summarized in Tab.~\ref{tab:transportcoefficients}. The Einstein relation
$\sigma_{m}=2eD_{m}\partial\rho_{m}/\hbar\partial\mu_{m}$ connects the magnon
diffusion constant $D_{m}$ defined by $\mathbf{j}_{m}=-D_{m}\boldsymbol{\nabla
}\rho_{m}$ with the magnon conductivity, where $\partial\rho_{m}/\partial
\mu_{m}=e\mathrm{Li}_{1/2}(e^{-\Delta/k_{B}T})/\left(  4\pi\Lambda
J_{s}\right)  $ and $\mathrm{Li}_{n}(z)$ is the poly-logarithmic function of
order $n$.

We observe that the magnon spin diffusion length $\ell_{\mathrm{mp}}$ is
smaller than the magnon decay length $\ell_{m}$ since the latter is
proportional to $\tau_{\mathrm{mr}}$, whereas $\ell_{\mathrm{mp}}$ is limited
by both magnon conserving and non-conserving scattering processes.
Furthermore, $1/\tau_{\mathrm{mr}}$ can be estimated by the
Landau-Lifshitz-Gilbert equation as $\sim\alpha_{G}k_{B}T/\hbar$
\cite{PhysRevB.90.094409}, where the Gilbert constant $\alpha_{G}$ at thermal
energies is not necessarily the same as for ferromagnetic resonance.

\begin{table}[ptb]
\begin{ruledtabular}
		\begin{tabular}{p{0.35\linewidth}cc}
			& Symbol & Expression  \Tstrut \\
			\hline
			Magnon thermal DeBroglie wavelength& $\Lambda$ & $\sqrt{4\pi J_s/\left( k_B T \right)}$ \Tstrut \\
			Magnon spin conductivity & $\sigma_m$ & $4 \zeta\left(3/2\right) ^2 e^2 J_s\tau/(\hbar^2 \Lambda^3)$ \\
			Magnon heat conductivity & $\kappa_m$ & $\frac{35}{2} \zeta\left(7/2\right) J_s k_B^2 T \tau / (\hbar^2 \Lambda^3 )$  \\
			Bulk spin Seebeck coefficient & $L$ & $10 \zeta\left(5/2\right) e J_s k_B T \tau /(\hbar^2 \Lambda^3 )$  \\
			Magnon thermal velocity & $v_{\rm th}$ & $2\sqrt{J_s k_B T} / \hbar$ \\
			Magnon spin diffusion length & $\ell_m$ & $v_{\rm th} \sqrt{\frac{2}{3}\tau \tau_{\rm mr}}$ \\
			Magnon-phonon relaxation length & $\ell_{\rm mp}$ & $v_{\rm th} \sqrt{\frac{2}{3}\tau \left(1/\tau_{\rm mr} + 1/\tau_{\rm mp} \right)^{-1}}$ \\
			Magnon spin-heat relaxation length & $\ell_{\rho T}$ & $\ell_m/\sqrt{\alpha_{\mu}}$ \\
			Magnon heat-spin relaxation length & $\ell_{Q\mu}$ & $\ell_m/\sqrt{\alpha_T}$ \\
			 & $\alpha_{\mu}$ & $\frac{5}{2} \zeta\left(5/2\right)/\zeta\left(3/2\right)$ \\
			 & $\alpha_T$ & $\frac{2}{7} \zeta\left(5/7\right)/\zeta\left(7/2\right)$ \\
			
		\end{tabular}
	\end{ruledtabular}
\caption{Transport coefficients and length scales \cite{2015arXiv150501329D}
as derived in Appendix~\ref{sec:appendix}. }%
\label{tab:transportcoefficients}%
\end{table}

\subsubsection{Clean systems}

In the limit of a clean system $1/\tau_{\mathrm{el}}\rightarrow0$. At
sufficiently low temperatures the magnon-conserving magnon-phonon scattering
rate $1/\tau_{\mathrm{mp}}\sim T^{3.5}$ \cite{PhysRev.152.731} (see also the
Appendix) loses against $1/\tau_{\mathrm{mr}}\sim\alpha_{G}k_{B}T/\hbar$ since
$\alpha_{G}$ is approximately temperature independent. Then all lengths
$\sim\Lambda/\alpha_{G}\sim10$\thinspace$\mathrm{\mu}$m for YIG at room
temperature and with $\alpha_{G}=10^{-4}$ from FMR \cite{PhysRevB.88.064408}.
The agreement with the observed signal decay \cite{Cornelissen2015} is likely
to be coincidental, however, since the spin waves at thermal energies have a
much shorter lifetime than the Kittel mode for which $\alpha_{G}$ is measured.
$\sigma_{m}$ estimated using the FMR Gilbert damping is larger than the
experimental value by several orders of magnitude, which is a strong
indication that the clean limit is not appropriate for realistic devices at
room temperature.

\subsubsection{Estimates for YIG at room temperature}

\label{subsubsec:estimatesforyig} The phonon and magnon inelastic mean free
paths derived from the experimental heat conductivity appear to be almost
identical at low temperatures up to 20 K \cite{PhysRevB.90.064421} but could
not be measured at higher temperatures. Both are likely to be limited by
the same scattering mechanism, i.e. the magnon-phonon interaction. We assume
here that the magnon-phonon scattering of thermal magnons at room temperature
is dominated by the exchange interaction (which always conserves magnons)
rather than the magnetic anisotropy (which may not conserve magnons)
\cite{PhysRevB.89.184413}. Then, $\tau\sim\tau_{\mathrm{mp}}$ and
extrapolating the low temperature results to room temperature leads to an
$\ell_{\mathrm{mp}}$ of the order of a nm, in agreement with an analysis of
spin Seebeck \cite{PhysRevB.88.094410} and Peltier \cite{Flipse2014}
experiments. The associated time scale $\tau_{\mathrm{mp}}\sim1-0.1$ ps is of
the same order as $\tau_{\mathrm{mm}}$ estimated in
Sec.~\ref{subsec:spintransportmagneticinsultors}. On the other hand,
$\tau_{\mathrm{mr}}\sim1$ ns from $\alpha_{G}\sim10^{-4}$ and therefore
$\ell_{m}\sim v_{\mathrm{th}}\sqrt{\tau_{\mathrm{mp}}\tau_{\mathrm{mr}}}%
\sim0.1-1$ $\mathrm{\mu}$m. The observed magnon spin transport signal decays
over a somewhat longer length scale ($\sim10$ $\mathrm{\mu}$m). Considering
that the estimated $\tau_{\mathrm{mr}}$ is an upper limit, our crude model
apparently overestimates the scattering. An important conclusion is,
nonetheless, that $\ell_{m}\gg\ell_{\mathrm{mp}}$, which implies that the
magnon chemical potential carries much farther than the magnon temperature.

With $\tau\sim\tau_{\mathrm{mp}}\sim0.1-1$ ps we can also estimate the magnon
spin conductivity $\sigma\sim e^{2}J_{s}\tau/\hbar^{2}\Lambda^{3}\sim
10^{5}-10^{6}$ S/m, in reasonable agreement with the value extracted from our
experiments (see next section).

\section{Heterostructures}

\label{subsec:applications}

Here we apply the model, introduced and parameterized in the previous section,
to concrete contact geometries and compare the results with experiments. We start
with an analytical treatment of the one-dimensional geometry, followed by
numerical results for the transverse configuration of top metal contacts on a
YIG film with finite thickness. Throughout, we assume ---motivated by the
estimates presented in the previous section--- that the magnon-phonon
relaxation is so efficient that the magnon temperature closely follows the
phonon temperature, i.e. $T_{m}=T_{p}$ (only in section
\ref{subsubsec:magnontemperature} we study the implications of the opposite
case, i.e. $T_{m}=T_{p}$ and $\mu_{m}=0$). This allows us to focus on the spin
diffusion equation for the chemical potential $\mu_{m}$. This approximation
should hold at room temperature, while the opposite regime $\ell_{\mathrm{mp}%
}\gg\ell_{m}$ might be relevant at low temperatures or high magnon densities:
when the magnon chemical potential is pinned to the band edge, transport can be
described in terms of the effective magnon temperature. The intermediate
regime $\ell_{\mathrm{mp}}\sim\ell_{m}$ in which both magnon chemical
potential and effective temperature have to be taken into account, is left for
future study. \begin{table}[ptb]
\begin{ruledtabular}
		\begin{tabular}{p{0.40\linewidth}ccc}
			& Symbol & Value & Unit  \Tstrut \\
			\hline
			YIG lattice constant & a & $12.376$ & $\text{\normalfont\AA}$ \Tstrut \\
			Spin quantum number per YIG unit cell & S & $10$ & - \\
			Spin wave stiffness constant in YIG & $J_s$ & $8.458\times10^{-40}$ & Jm$^{2}$ \\
			YIG magnon spin diffusion length & $\ell_m$ & $9.4$ & $\upmu$m \\
			YIG spin conductivity & $\sigma_m$ & $5 \times 10^5$ & S/m\\ 
			Real part of the spin mixing conductance & $g^{\uparrow\downarrow}$ & $1.6\times10^{14}$ & S/m$^2$ \\ 
			Platinum conductivity & $\sigma_e$ & $2.0\times10^6$ & S/m \\
			Platinum spin relaxation length & $\ell_s$ & $1.5$ & nm\\
			Platinum spin Hall angle & $\theta$ & $0.11$ & -
		\end{tabular}
	\end{ruledtabular}
\caption{Selected parameters for spin and heat transport in bilayers with
magnetic insulators and metals. $\mathrm{a}$, $\mathrm{S}$ and $J_{s}$ are
adopted from \cite{Cherepanov1993}, $\ell_{s}$ and $\theta$ from
\cite{Flipse2014,Weiler2013}, and $\sigma_{e}$ is extracted from electrical
measurements on our devices \cite{Cornelissen2015}. Note that our values for $\sigma_e$ and $\ell_s$ are consistent with Elliot-Yafet scattering as the dominant spin relaxation mechanism in platinum \cite{Nguyen2016}. The mixing conductance,
magnon spin diffusion length, and the magnon spin conductivity are estimated
in the main text.}%
\label{tab:modelparameters}%
\end{table}

\subsection{One-dimensional model}

\label{subsec:onedimension} We consider first the one-dimensional geometry
shown in Fig.~\ref{fig:geometry1D}. We focus on strictly linear response and
therefore disregard Joule heating in the metal contacts as well as
thermoelectric voltages by the spin Nernst and Ettingshausen effects. The spin
and charge currents in the metal are then governed by
\begin{equation}%
\begin{pmatrix}
j_{c}\\
\frac{2e}{\hbar}j_{s}%
\end{pmatrix}
=%
\begin{pmatrix}
\sigma_{e} & -\sigma_{\mathrm{SH}}\\
-\sigma_{\mathrm{SH}} & -\sigma_{e}%
\end{pmatrix}%
\begin{pmatrix}
\partial_{y}\mu_{e}\\
\frac{1}{2}\partial_{x}\mu_{z}%
\end{pmatrix},
\label{eq:spindrifteqsoned}%
\end{equation}
where the charge transport is in the $y$-direction, spin transport in the
$x$-direction, and the electron spin accumulation is pointing in the
$z$-direction. The spin and magnon diffusion equations reduce to
\begin{align}
\frac{\partial^{2}\mu_{s}}{\partial x^{2}}  &  =\frac{\mu_{z}}{\ell_{s}^{2}%
},\label{eq:spintransport1D}\\
\frac{\partial^{2}\mu_{m}}{\partial x^{2}}  &  =\frac{\mu_{m}}{\ell_{m}^{2}}~.
\label{eq:magnontransport1D}%
\end{align}
The interface spin currents Eq.~(\ref{eq:spincurrentinterface}) provide the
boundary conditions at the interface to the ferromagnet, while all currents at
the vacuum interface vanish. Eqs.~(\ref{eq:interfacecurrents}) and
(\ref{eqn:gs}) lead to the interface spin current density $j_{s}%
^{\mathrm{int}}=g_{s}\left(  \mu_{z}^{\mathrm{int}}-\mu_{m}^{\mathrm{int}%
}\right)  $, where $g_{s}$ is defined in Eq.~(\ref{eqn:gs}).

\subsubsection{Current transfer efficiency}

\label{subsec:transferefficiency} 
\begin{figure}[h]
\includegraphics[width=8.5cm]{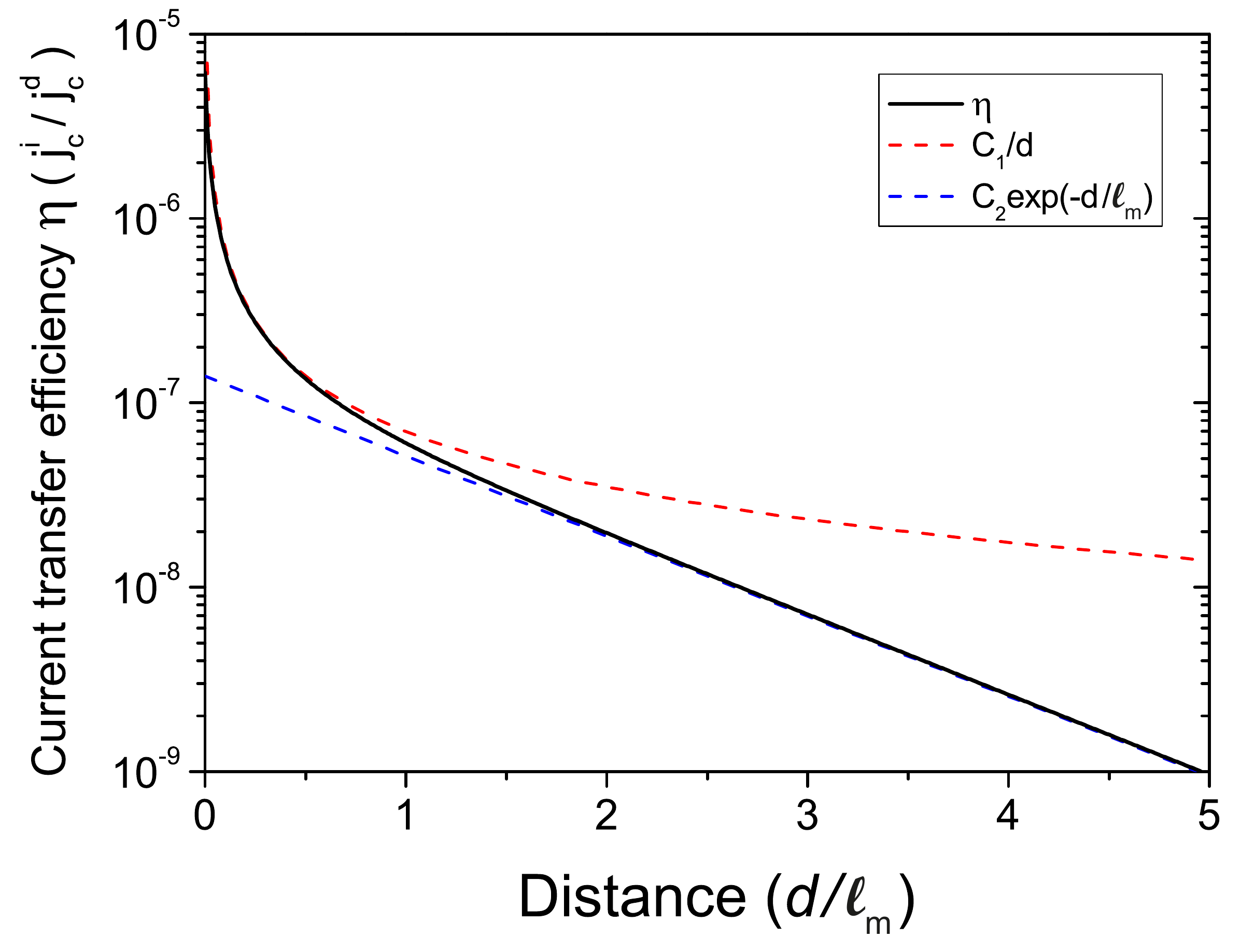}\caption{The
current transfer efficiency $\eta$ (non-local resistance normalized by that of
the metal contacts) as a function of distance between the contacts in a
Pt$|$YIG$|$Pt structure calculated in the 1D model. Parameters are taken from
Tab.~\ref{tab:modelparameters} and the Pt thickness $t=10$ nm. The dashed
lines are plots of the functions $C_{1}/d$ (red dashed line) and $C_{2}%
\exp\left(  -d/\ell_{m}\right)  $ (blue dashed line) to show the different
modes of signal decay in different regimes: diffusive $1/d$ decay for
$d<\ell_{m}$ and exponential decay for $d>\ell_{m}$. The constants $C_{1}$ and
$C_{2}$ were chosen to show overlap with $\eta$ for illustrative purposes, but
have no physical meaning.}%
\label{fig:etavsL}%
\end{figure}

The non-local resistance $R_{\mathrm{nl}}$
is the voltage over the detector divided by current in the injector, also
referred to as non-local spin Hall magnetoresistance (see below). The magnon
spin injection and detection can also be expressed in terms of the current
transfer efficiency $\eta$, i.e. the absolute value of the ratio between the
currents in the detector and injector strip \cite{PhysRevLett.109.096603} when
the detector circuit is shorted. $\eta=R_{\mathrm{nl}}/R_{0}$ for identical Pt
contacts with resistance $R_{0}$. In Fig.~\ref{fig:etavsL} we plot the
calculated $\eta$ as a function of distance $d$ between the contacts for a Pt
thickness $t=10$ nm and parameters from Table~\ref{tab:modelparameters}.
$\eta$ decays algebraically $\propto1/d$ when $d\ll\ell_{m},$ which implies
diffusion without relaxation, and exponentially for $d\gg\ell_{m}$. The
calculated order of magnitude already agrees with experiments
\cite{Cornelissen2015}. The $\eta^{\prime}$s in
Ref.~[\onlinecite{PhysRevLett.109.096603}] are three orders of magnitude
larger than ours due to their much weaker relaxation.

The origin of the small $\eta$ is caused by the inefficiency of the spin-Hall
mediated spin-charge conversion. The ratio between the spin accumulations in
injector and detector $\eta_{s}=\mu_{s}^{\mathrm{det}}/\mu_{s}^{\mathrm{inj}}$
is much larger than $\eta$ and discussed in
Sec.~\ref{subsubsec:spintransferefficiency}.

\subsubsection{Spin Hall magnetoresistance}

\label{subsubsec:smr} 
\begin{figure}[h]
\includegraphics[width=8.5cm]{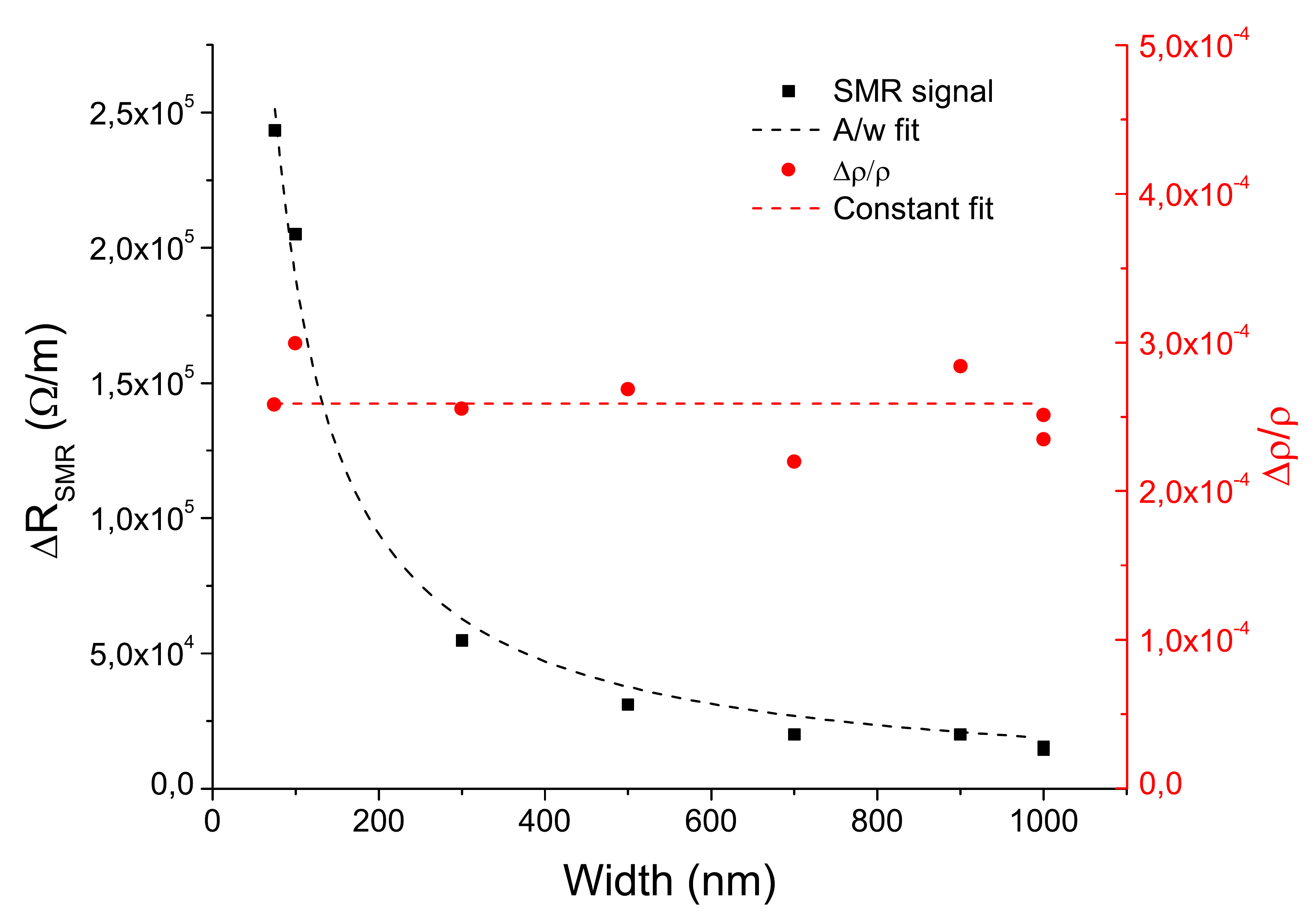} \caption{Experimental
spin Hall magnetoresistance (SMR) as a function of platinum strip width. The
black squares (left axis) show absolute resistance changes $\Delta
R_{\mathrm{SMR}}$ devided by the device length (18 $\mathrm{\mu}$m) in units
of $\mathrm{\Omega}$/m . The red dots (right axis) show the relative
resistivity changes $\Delta\rho/\rho$. }%
\label{fig:smr}%
\end{figure}

The effective spin mixing conductance $g_{s}$ governs
the amount of spin transferred across the interface between the normal metal
and the magnetic insulator. While $g_{s}$ cannot be extracted from
measurements directly, it is related to the spin mixing conductance
$g^{\uparrow\downarrow}$ via Eq.~(\ref{eqn:gs}). In order to determine
$g^{\uparrow\downarrow}$ we measured the spin Hall magnetoresistance (SMR)
\cite{Nakayama2013,Vlietstra2013a} in devices of Ref.~\cite{Cornelissen2015}.
The SMR is defined as the relative resistivity change in the Pt contact between
in-plane magnetization parallel and normal to the current, $\Delta\rho/\rho$. The expression for the magnitude of the SMR reads \cite{Chen2013}
\begin{equation}
\frac{\Delta\rho}{\rho}=\theta^{2}\frac{\ell_{s}}{t}\frac{2\ell_{s}%
g^{\uparrow\downarrow}\tanh^{2}\frac{t}{2\ell_{s}}}{\sigma_{e}+2\ell
_{s}g^{\uparrow\downarrow}\coth\frac{t}{\ell_{s}}}, \label{eq:deltarhooverrho}%
\end{equation}
where $t=13.5$ nm is the platinum thickness. Figure~\ref{fig:smr} shows the
experimental SMR as a function of platinum strip width. As expected
$\Delta\rho/\rho=(2.6\pm0.09)\times10^{-4}$ does not depend on the strip
width. Using Eq.~(\ref{eq:deltarhooverrho}) and the values for $\ell_{s}$,
$\theta$ and $\sigma_{e}$ as indicated in Tab.~\ref{tab:modelparameters}, we
find $g^{\uparrow\downarrow}=(1.6\pm0.06)\times10^{14}$ S/m$^{2},$which agrees
with previous reports \cite{Vlietstra2013a,Jungfleisch2013,Weiler2013}.

In Chen \textit{et al.}'s zero-temperature theory \cite{Chen2013} the spin
current generated by the spin Hall effect in Pt is perfectly reflected when
spin accumulation and magnetization are collinear. As discussed above, at
finite temperature a fraction of the spin current is injected into the
ferromagnet in the form of magnons. This implies that the SMR should be a
monotonously decreasing function of temperature. This has been found for high
temperatures \cite{Uchida2015}, but the decrease of the SMR at low
temperatures \cite{Meyer2014} hints at a temperature dependence of other
parameters such as the spin Hall angle.

The current transfer efficiency $\eta$ can be interpreted as a non-local
version of the SMR \cite{Goennenwein2015} The SMR is caused by the contrast in
spin current absorption of the YIG%
$\vert$%
{}Pt interface when the spin accumulation vector is normal or parallel to the
magnetization $\vec{M}$. In the non-local geometry, we measure the voltage in
contact 2 that has been induced by a charge current (in the same direction) in
contact 1. Since $g_{s}<g^{\uparrow\downarrow}$ the relation $\left\vert
\Delta\rho/\rho\right\vert \geq\eta$ must hold even in the absence of losses
in the ferromagnet and detector. This indeed agrees with our data.

\subsubsection{Interface transparency}

The analytical expression for $\eta$ in the one-dimensional geometry is
lengthy and omitted here, but it can be simplified for special cases. In the
the limit of a large bulk magnon spin resistance, the interface resistance can
be disregarded. The decay of the spin current is then dominated by the bulk
spin resistance and relaxation of both materials. When $\sigma_{m}/\ell
_{m},\sigma_{e}/\ell_{s}\ll g_{s}$
\begin{equation}
\eta=\frac{\theta^{2}\ell_{m}\sigma_{e}\sigma_{m}}{t\left[  \sigma_{m}%
^{2}+\left(  \frac{\ell_{m}}{\ell_{s}}\right)  ^{2}\sigma_{e}^{2}\right]
}\sinh^{-1}\frac{d}{\ell_{m}}~, \label{eq:etatransparantinterface}%
\end{equation}
where the Pt thickness is chosen $t\gg\ell_{s}$ and $\theta=\sigma
_{\mathrm{SH}}/\sigma_{e}$ is the spin Hall angle. When $d\ll\ell_{m}$ we are
in the purely diffusive regime with algebraic decay $\eta\propto1/d$.
Exponential decay with characteristic length $\ell_{m}$ takes over when
$d\gtrsim\ell_{m}$. In our experiments (see Tab.~\ref{tab:modelparameters})
$\sigma_{m}\sim\sigma_{e}$ and $\ell_{m}\gg\ell_{s}$, so
\begin{equation}
\eta=\frac{\theta^{2}\ell_{s}^{2}\sigma_{m}}{\ell_{m}t\sigma_{e}}\sinh
^{-1}\frac{d}{\ell_{m}}~. \label{eq:etatransparantinterface2}%
\end{equation}

On the other hand, when $\sigma_{m}/\ell_{m},\sigma_{e}/\ell_{s}\gg g_{s}$ the
interfaces dominate and
\begin{equation}
\eta=\frac{\theta^{2}g_{s}^{2}\ell_{s}^{2}\ell_{m}}{t\sigma_{e}\sigma_{m}%
}\sinh^{-1}\frac{d}{\ell_{m}}~,
\end{equation}
with identical scaling with respect to $d$, but a different prefactor.
According to the parameters in Tab.~\ref{tab:modelparameters} $\sigma_{m}%
/\ell_{m}\gg\sigma_{e}/\ell_{s}\gg g_{s}$, so spin injection is limited by the
interfaces due to the small spin conductance between YIG and platinum.

\subsection{Two-dimensional geometry}

\label{subsec:twodimensions} 
\begin{figure*}[ptb]
\includegraphics[width=18cm]{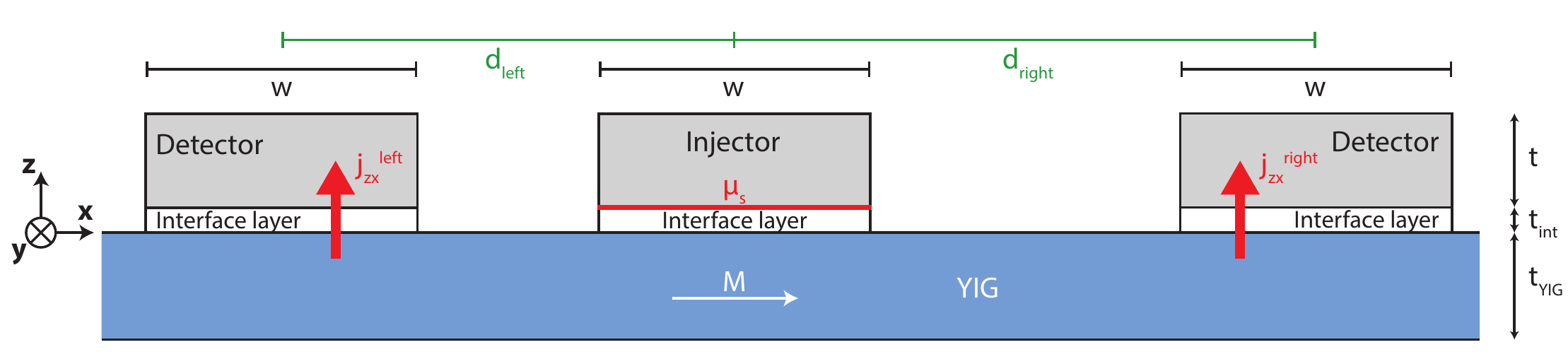} \caption{Schematic of the 2D geometry. The relevant dimensions are indicated in the figure. The spin accumulation arising from the charge current through the
injector, $\mu_{s}$, is used as a boundary condition on the YIG$|${}Pt
interface. The interface layer is used to account for the effect of finite
spin mixing conductance between YIG and platinum. }%
\label{fig:geometry_2D}%
\end{figure*}

Experiments are carried out for Pt%
$\vert$%
YIG%
$\vert$%
Pt with a lateral (transverse) geometry in which the platinum injector and
detector are deposited on a YIG film. The two-dimensional model sketched in
Fig.~\ref{fig:geometry_2D} captures this configuration but cannot be treated
analytically. We therefore developed a finite-element implementation of our
spin diffusion theory by the COMSOL Multiphysics (version 4.3a) software
package, extending the description of spin transport in metallic
systems~\cite{Slachter2011a} to magnetic insulators. The finite-element simulations of the spin Seebeck
\cite{PhysRevB.88.094410} and spin Peltier~\cite{Flipse2014} effects in Pt%
$\vert$%
YIG focussed on heat transport and were based on a magnon temperature
diffusion model. Here we find that neglecting the magnon chemical potential
underestimates spin transport by orders of magnitude, because the magnon
temperature equilibrates at a length scale $\ell_{\mathrm{mp}}$ of a few
nanometers and the magnon heat capacity and heat conductivity are small
\cite{PhysRevB.90.064421}. The magnon chemical potential and the associated
non-equilibrium magnons, on the other hand, diffuse on the much longer length
scale $\ell_{m}$.

In order to model the experiments in two dimensions, we assume translational
invariance in the third direction, which is justified by the large aspect
ratio of relatively small contact distances compared with their length. With
equal magnon and phonon temperatures everywhere, the magnon transport in two
dimensions is governed by
\begin{align}
\frac{2e}{\hbar}\mathbf{j}_{m}  &  =-\sigma_{m}\boldsymbol{\nabla}\mu
_{m}~,\nonumber\\
\nabla^{2}\mu_{m}  &  =\frac{\mu_{m}}{\ell_{m}^{2}}~,
\end{align}
where $\boldsymbol{\nabla}=\mathbf{x}\partial_{x}+\mathbf{z}\partial_{z}$.

The particle spin current $\mathbf{j}_{s}=(j_{\mathrm{xx}},j_{\mathrm{zx}})$
in the metal is described by
\begin{align}
\frac{2e}{\hbar}\mathbf{j}_{s}  &  =-\frac{\sigma_{e}}{2}\boldsymbol{\nabla
}\mu_{x}~,\nonumber\\
\nabla^{2}\mu_{x}  &  =\frac{\mu_{x}}{\ell_{s}^{2}}~,
\end{align}
where $\mu_{x}$ is the $x$-component of the electron spin accumulation. The
spin-charge coupling via the spin Hall effect is implemented by the boundary
conditions in Sec.~\ref{subsubsec:boundaryconditions}, while the inverse spin
Hall effect is accounted for in the calculation of the detector voltage, see
Sec.~\ref{subsubsec:nonlocalresistance}). The estimates at the end of the
previous section justify disregarding temperature effects.

\subsubsection{Geometry}

\label{subsubsec:2Dgeometry} In order to accurately model the experiments, we
define two detectors (left and right) and a central injector, introducing the
distances $d_{\mathrm{left}}$ and $d_{\mathrm{right}}$ as in
Fig.~\ref{fig:geometry_2D}. We generate a short (A) and a long distance (B)
geometry. The injector and detectors are slightly different as summarized in
table \ref{tab:geometry_params}. The YIG film thicknesses are $200$ nm for (A)
and $210$ nm for (B). \begin{table}[h]
\begin{ruledtabular}
		\begin{tabular}{ccccc}
			& Pt width & Pt thickness & Distances  \Tstrut \\
			& $w$ (nm) & $t$ (nm) & $d$ ($\upmu$m) \\
			\hline
			Geometry A & $140$ & $13.5$ & $0.2-5$ \Tstrut\\
			Geometry B & $300$ & $7$ & $2-42.5$			
		\end{tabular}
	\end{ruledtabular}
\caption{Properties of geometry sets A and B.}%
\label{tab:geometry_params}%
\end{table}The YIG film is chosen to be long compared to the spin diffusion
length ($w_{\mathrm{YIG}}=150$ $\mathrm{\mu}$m) in order to prevent
finite-size artifacts.

\subsubsection{Boundary conditions}

\label{subsubsec:boundaryconditions} Sending a charge current density $j_{c}$
in the $+y$-direction through the platinum injector strip generates a spin
accumulation $\mu_{s}$ at the YIG%
$\vert$%
{}platinum interface by the spin Hall effect (shown in
Fig.~\ref{fig:geometry_2D}). This is captured by
Eqs.~(\ref{eq:spindrifteqsmetal}) that predict a spin accumulation at the Pt
side of the interface of \cite{Flipse2014}
\begin{equation}
\mu_{s}\equiv\left.  \mu_{x}\right\vert _{\mathrm{interface}}=2\theta
j_{c}\frac{\ell_{s}}{\sigma_{e}}\tanh\left(  \frac{t}{2\ell_{s}}\right)  ,
\label{eqn:boundary_mus}%
\end{equation}
which is used for the interface boundary condition of the magnon diffusion
equation.
Here, we assume that the contact with the YIG does not significantly affect
the spin accumulation \cite{Chen2013}, which is allowed for the collinear
configuration since $g_{s}<\sigma_{e}/\ell_{s}$. The spin orientation of
$\mu_{s}$ points along $-\mathbf{x}$, parallel to the YIG magnetization.
A charge current $I=100$~$\mathrm{\mu}$A generates spin accumulations in the
injector contact of $\mu_{s}^{A}=9.6$~$\mathrm{\mu}$V and $\mu_{s}^{B}%
=7.7$~$\mathrm{\mu}$V for geometries A and B, respectively.

The uncovered YIG surface is subject to a zero current boundary condition
$\left(  \boldsymbol{\nabla}\cdot\mathbf{n}\right)  \boldsymbol{\mu}%
_{s}=\mathbf{0}$, where $\mathbf{n}$ is the surface normal.
\begin{figure}[ptb]
\includegraphics[width=8.5cm]{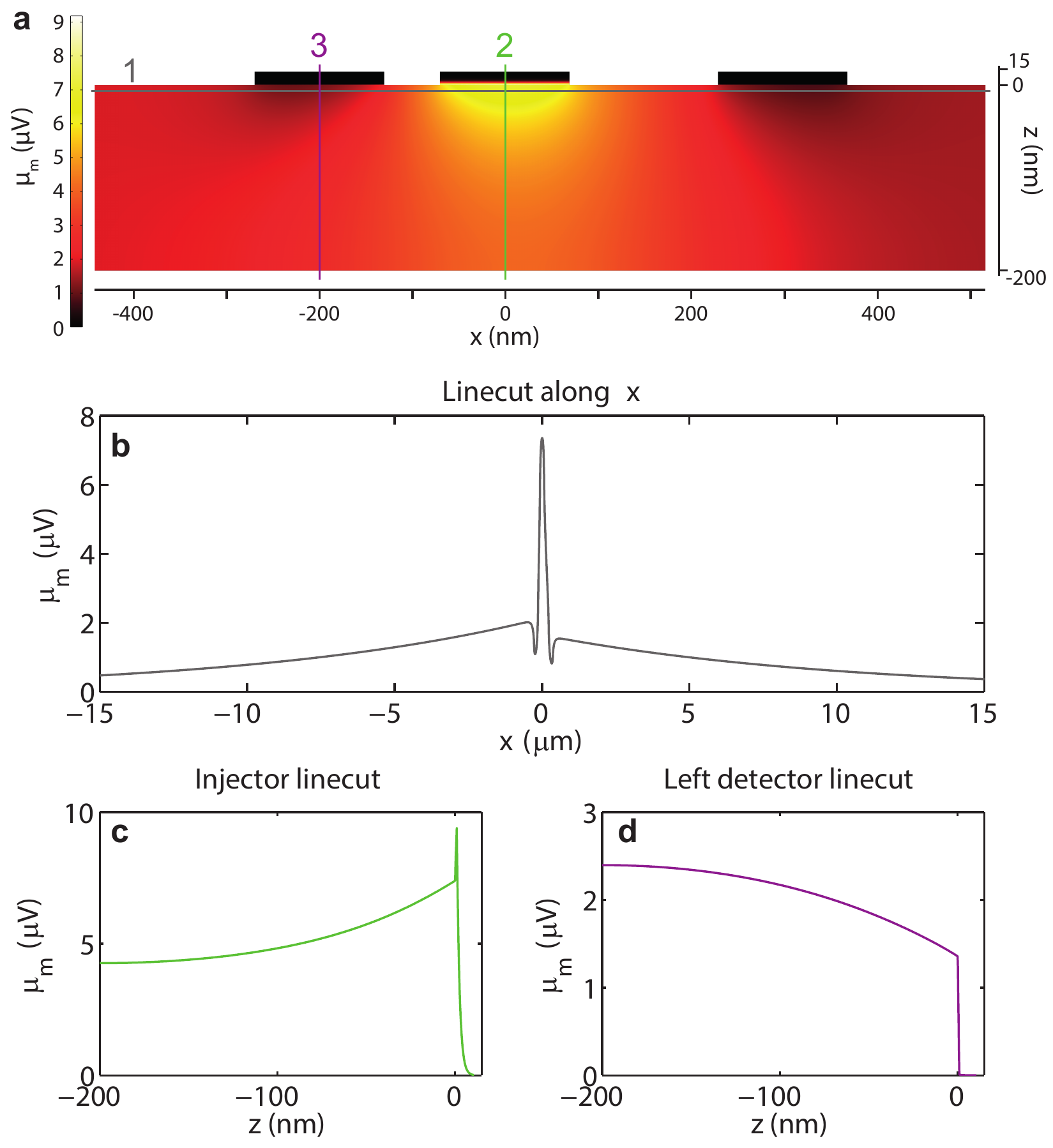} \caption{(a) Two-dimensional magnon chemical potential distribution for geometry (A) with
$d_{\mathrm{left}}=200$ nm and $d_{\mathrm{right}}=300$ nm. The lines numbered
1,2,3 indicate the locations of the profiles plotted in figures (b),(c),(d),
respectively. }%
\label{fig:muprofiles}%
\end{figure}\begin{figure*}[ptb]
\includegraphics[width=18cm]{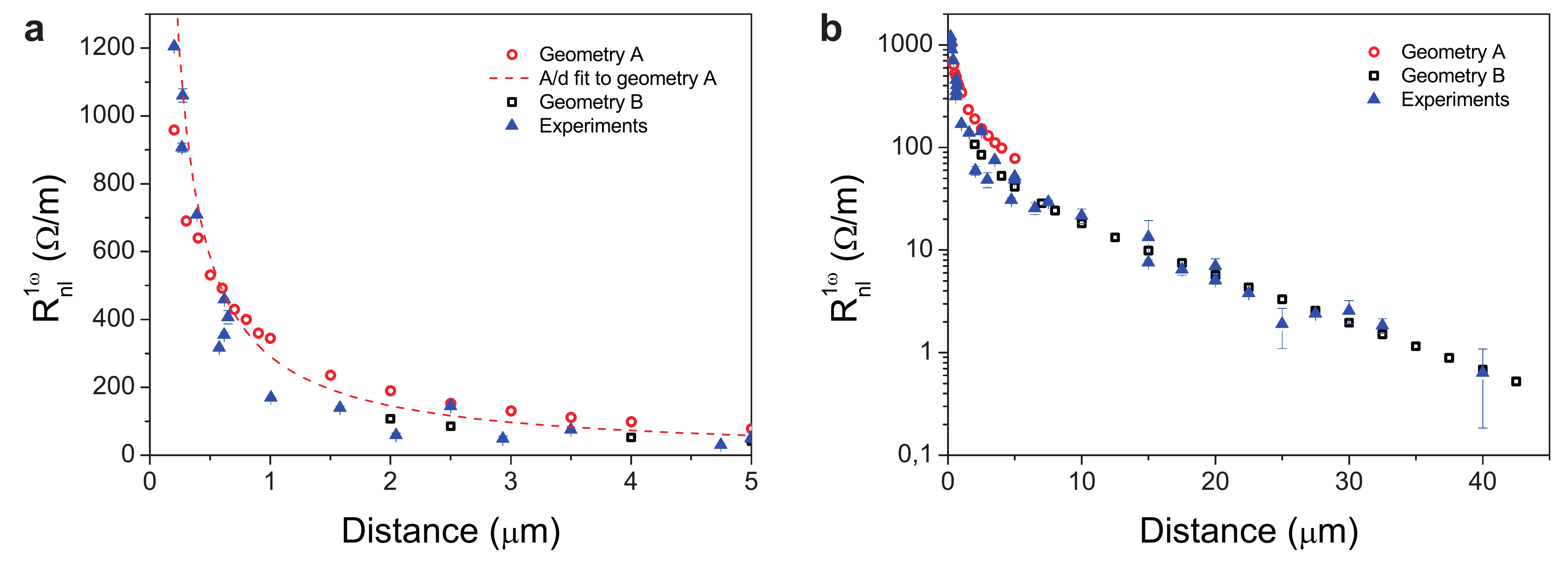} \caption{(a) Computed non-local first harmonic signal as a function of distance on a linear
scale. The red open circles show the results for sample (A), while black open
squares represent sample (B). The blue triangles are the experimental results
\cite{Cornelissen2015}. The red dashed line is a $1/d$ fit of the numerical
results for (A). (b) Same as (a) but on a logarithmic scale.}%
\label{fig:results2D}%
\end{figure*}

\subsubsection{The YIG%
$\vert$%
Pt interface}

\label{subsubsec:interface}The interface spin conductance $g_{s}$ is modelled
by a thin interface layer, leading to a spin current $j_{s}^{\mathrm{int}%
}=-\sigma_{s}^{\mathrm{int}}\partial\mu_{x}/\partial z$, with spin
conductivity $\sigma_{s}^{\mathrm{int}}=g_{s}t_{\mathrm{int}}$. When the
interface thickness $t_{\mathrm{int}}$ is small compared to the platinum
thickness $t_{\mathrm{Pt}}$ we can accurately model the Pt%
$\vert$%
YIG interface without having to change the COMSOL code. Varying the auxiliary
interface layer thickness between $0.5<t_{\mathrm{int}}<2.5$~nm, the spin
currents vary by only $0.1$\%. In the following we adopt $t_{\mathrm{int}%
}=1.0$~nm.

Finally, with Eq.~(\ref{eqn:gs}) $g_{s}=0.06g^{\uparrow\downarrow}$ and
$g^{\uparrow\downarrow}$ from Sec.~\ref{subsubsec:smr} we get $g_{s}%
=9.6\times10^{12}$ S/m$^{2}$.

\subsubsection{Magnon chemical potential profile}

\label{subsubsec:muprofile} A representative computed magnon chemical
potential map is shown in Fig.~\ref{fig:muprofiles}(a), while different
profiles along the three indicated cuts are plotted in
Fig.~\ref{fig:muprofiles}(b)-(d). The magnon chemical potential along $x$ and
at $z=-1$ nm (i.e. 1 nm below the surface of the YIG) in
Fig.~\ref{fig:muprofiles}(b) is characterized by the spin injection by the
center electrode. Globally, $\mu_{m}$ decays exponentially with distance from
the injector on the scale of $\ell_{m}$. We also observe that the left and
right detector contacts at $x=-200$ nm and $x=300$ nm, respectively, act as
sinks that visibly suppress but do not quench the magnon accumulation. The
finite mixing conductance and therefore magnon absorption are also evident
from the profiles along $z$ in Figs.~\ref{fig:muprofiles}(c) and
\ref{fig:muprofiles}(d): The magnon chemical potential changes abruptly across
the YIG%
$\vert$%
{}Pt interface by the relatively large interface resistance $g_{s}^{-1}$. The
magnon chemical potential is much smaller than the magnon gap ($\sim1$ K). We
are therefore far from the threshold for current-driven instabilities such as
magnon condensation and/or self-oscillations of the magnetization
\cite{PhysRevB.90.094409}.

\subsubsection{Detector contact and non-local resistance}

\label{subsubsec:nonlocalresistance} The spin current density in the detectors
is governed by the spin accumulation according to
\begin{equation}
\left\langle j_{zx}\right\rangle =-\frac{\sigma_{e}}{2A}\int_{A}\frac
{\partial\mu_{x}}{\partial z}dA^{\prime}, \label{eqn:detectorspincurrent}%
\end{equation}
which is an average over the detector area $A=wt$. The observable non-local
resistance $R_{\mathrm{nl}}$ (normalized to device length) in units of
$\mathrm{\Omega}$/m
\begin{equation}
R_{\mathrm{nl}}=\frac{\theta\left\langle j_{zx}\right\rangle }{\sigma_{e}I}.
\label{eqn:nonlocalresistance}%
\end{equation}
is compared with experiments in the next section.

\subsection{Comparison with experiments}

\label{subsec:comparison}

\subsubsection{Two-dimensional model}

\label{subsubsec:FEM} Fig.~\ref{fig:results2D} compares the simulations as
described in the previous section with our experiments \cite{Cornelissen2015}.
Fig.~\ref{fig:results2D}(a) is a linear plot for closely spaced Pt contacts
while Fig.~\ref{fig:results2D}(b) shows the results for all contact distances
on a logarithmic scale. The magnon spin conductivity $\sigma_{m}$ and the
magnon spin diffusion length $\ell_{m}$ are adjustable parameters; all others
are listed in Table.~\ref{tab:modelparameters}. We adopted $\sigma_{m}%
=5\times10^{5}$~S/m and $\ell_{m}=9.4$ $\mathrm{\mu}$m as the best fit values
that agree with the estimates in Ref.~\cite{Cornelissen2015} and
Sec.~\ref{subsec:lengthscales}.

At large contact separations in geometry (B), the signal is more sensitive to
the bulk parameters $\ell_{m}$ and $\sigma_{m}$ than the interface $g_{s}$.
When contacts are close to each other, the interfaces become more important
and the results depend sensitively on $g_{s}$ and $\sigma_{m}$ as compared to
$\ell_{m}$. For very close contacts ($d<500$~nm) the total spin resistance of
YIG is dominated by the interface and our model calculations slightly
underestimate the experimental signal and, in contrast to experiments, deviate
from the $\sim d^{-1}$ fit that might indicate an underestimated $g_{s}.$
However, a larger $g_{s}$ would lead to deviations at intermediate distances
($1<d<5$~$\mathrm{\mu}$m).

\subsubsection{Spin transfer efficiency and equivalent circuit model}

\label{subsubsec:spintransferefficiency} The spin transfer efficiency
$\eta_{s}=\mu_{s}^{\mathrm{det}}/\mu_{s}^{\mathrm{inj}}$, i.e. the ratio
between the spin accumulation in the injector and that in the detector, can be
readily derived from the experiments by Eq.~(\ref{eqn:boundary_mus}). From the
voltage generated in the detector by the inverse spin Hall effect
$V_{\mathrm{ISHE}}$ \cite{Castel2012}
\begin{equation}
\mu_{s}^{\mathrm{det}}=\frac{2t}{\theta L}\frac{1+e^{-2t/\ell_{s}}}{\left(
1-e^{-t/\ell_{s}}\right)  ^{2}}V_{\mathrm{ISHE}},
\label{eqn:Castelspinaccumulation}%
\end{equation}
where $l$ 
is the length of the metal contact. The
spin transfer efficiency therefore reads
\begin{equation}
\eta_{s}=\frac{t}{\ell_{s}\theta^{2}}\frac{R_{\mathrm{nl}}}{R_{\mathrm{det}}%
}\frac{\left(  e^{t/\ell_{s}}+1\right)  \left(  e^{2t/\ell_{s}}+1\right)
}{\left(  e^{t/\ell_{s}}-1\right)  ^{3}}, \label{eq:etas}%
\end{equation}
where $R_{\mathrm{nl}}=V_{\mathrm{ISHE}}/I$ is the observed non-local
resistance and $R_{\mathrm{det}}$ the detector resistance. \begin{figure}[ptb]
\includegraphics[width=8.5cm]{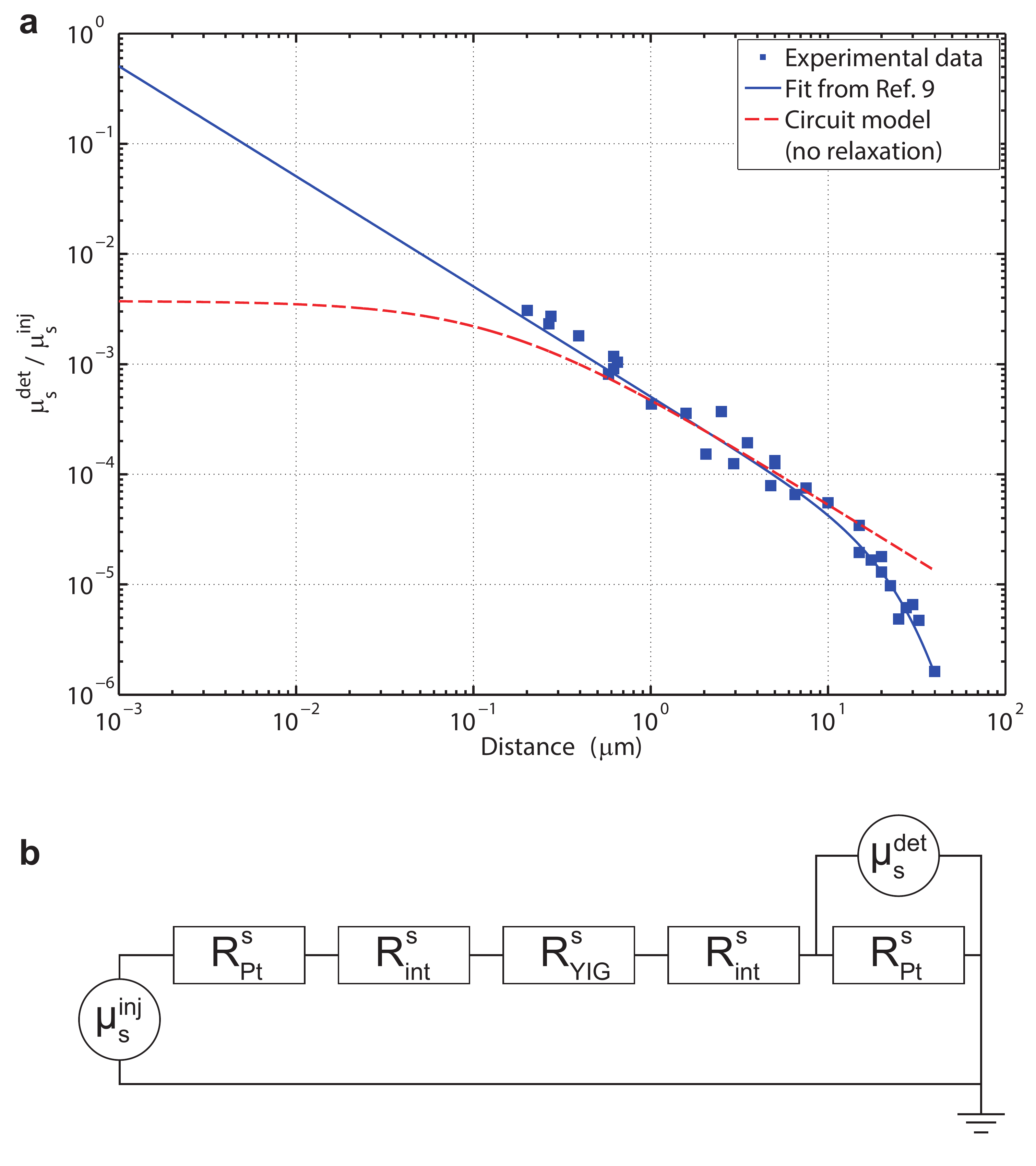} \caption{(Color
online) (a) Experimental and simulated spin transfer efficiency $\eta_{s}%
=\mu_{s}^{\mathrm{det}}/\mu_{s}^{\mathrm{inj}}$. The blue solid line is a fit
by the 1D spin diffusion model \cite{Cornelissen2015}. Since here interfaces
are disregarded $\mu_{s}^{\mathrm{det}}\rightarrow\mu_{s}^{\mathrm{inj}}$ for
vanishing contact distances. The red dashed line are obtained from the
equivalent circuit model in (b) with spin resistances $R_{X}^{s}$ defined in
the text. This model includes $g_{s}$ but is valid for $d<\ell_{m}$ only since
spin relaxation is disregared. The interfaces lead to a saturation of
$\eta_{s}$ at short distances.}%
\label{fig:spin_transfer_efficiency}%
\end{figure}Figure~\ref{fig:spin_transfer_efficiency}a shows the experimental
data converted to the spin transfer efficiency as a function of distance $d$
that is fitted to a 1D magnon spin diffusion model that does not include the
interfaces \cite{Cornelissen2015}. When $d\rightarrow0$ and interfaces are
disregarded, $\eta_{s}$ diverges. This artifact can be repaired by the
equivalent spin-resistor circuit in Fig.~\ref{fig:spin_transfer_efficiency}(b)
according to which
\begin{equation}
\eta_{s}=\frac{R_{\mathrm{Pt}}^{s}}{R_{\mathrm{YIG}}^{s}+2R_{\mathrm{int}}%
^{s}+2R_{s}^{\mathrm{Pt}}}, \label{eq:circuit}%
\end{equation}
where $R_{\mathrm{Pt}}^{s}=\ell_{s}/\left(  \sigma A_{\mathrm{int}}%
\tanh(t/\ell_{s})\right)  $ is the spin resistance of the platinum strip
\cite{Castel2012}, $R_{\mathrm{int}}^{s}=1/(g_{s}A_{\mathrm{int}})$ is
interface spin resistance and $R_{\mathrm{YIG}}^{s}=d/\left(  \sigma
_{m}A_{\mathrm{YIG}}\right)  $ is the magnonic spin resistance of YIG.
$A_{\mathrm{YIG}}=lt_{\mathrm{YIG}}$ is the cross-section of the YIG channel
and $A_{\mathrm{int}}=wl$ is the area of the Pt%
$\vert$%
YIG interfaces. The parameters in Tab.~\ref{tab:modelparameters} lead to the
red dashed line in Fig.~\ref{fig:spin_transfer_efficiency}(a), which agrees
well with the experimental data for $d<\ell_{m}$. No free parameters were used
in this model, since we adopted $\sigma_{m}=5\times10^{5}$ S/m as extracted
from our 2D model in the previous section.

The model predicts that the spin transfer efficiency should saturate for
$d\lesssim100$ nm for $g_{s}=9.6\times10^{12}$ S/m$^{2}$. A predicted onset of
saturation at 200 nm is not confirmed by the experiments, which as pointed out
already in the previous section, could imply a larger $g_{s}$. Experiments on
samples with even closer contacts are difficult but desirable. Based on the
available data we predict that the efficiency saturates at $\eta_{s}%
=4\times10^{-3}$. The charge transfer efficiency (defined in Sec.\thinspace
\ref{subsec:transferefficiency}) would be maximized at $\eta\approx
5\times10^{-5}$, which is still below the SMR $\Delta\rho/\rho=2.6\times
10^{-4}$, as predicted in Sec.\thinspace\ref{subsubsec:smr}.

\subsubsection{Magnon temperature model}

\label{subsubsec:magnontemperature} We can analyze the experiments also in
terms of magnon temperature diffusion \cite{Sanders1977} as applied to the
spin Seebeck \cite{Xiao2010,PhysRevB.88.094410} and spin Peltier
\cite{Flipse2014} effects. Communication between the platinum injector and
detector is possible via phonon and magnon heat transport: The spin
accumulation at the injector can heat or cool the magnon/phonon system by the
spin Peltier effect. The diffusive heat current generates a voltage at the
detector by the spin Seebeck effect. However, pure phononic heat transport
does not stroke with the exponential scaling, but decays only logarithmically
(see below). The magnon temperature model (which describes the magnons in terms of their temperature only) can give an exponential scaling,
but in order to agree with experiments, the magnon-phonon relaxation length
must be large such that $T_m\neq T_p$ over large distances. This is at odds with the analysis by Schreier \emph{et al.}
and Flipse \emph{et al.}. However, we can test this model by, for the sake of
argument, increasing this length scale by four orders of magnitude to:
$\ell_{\mathrm{mp}}=9.4$ $\mathrm{\mu}$m and completely disregard the magnon
chemical potential. The spin Peltier heat current $Q_{\mathrm{SPE}%
}^{\mathrm{inj}}$ is then \cite{Flipse2014}
\begin{equation}
Q_{\mathrm{SPE}}^{\mathrm{inj}}=L_{s}T\frac{\mu_{s}^{\mathrm{inj}}}%
{2}A^{\mathrm{int}},
\end{equation}
where $L_{s}$ is the interface spin Seebeck coefficient, $L_{s}=2g^{\uparrow
\downarrow}\gamma\hbar k_{B}/(eM_{s}\Lambda^{3})$
\cite{Xiao2010,Flipse2014,PhysRevB.88.094410}, and $M_{s}=\mu_B S/a^3$ is
the saturation magnetization of YIG. \begin{figure}[ptb]
\includegraphics[width=8.5cm]{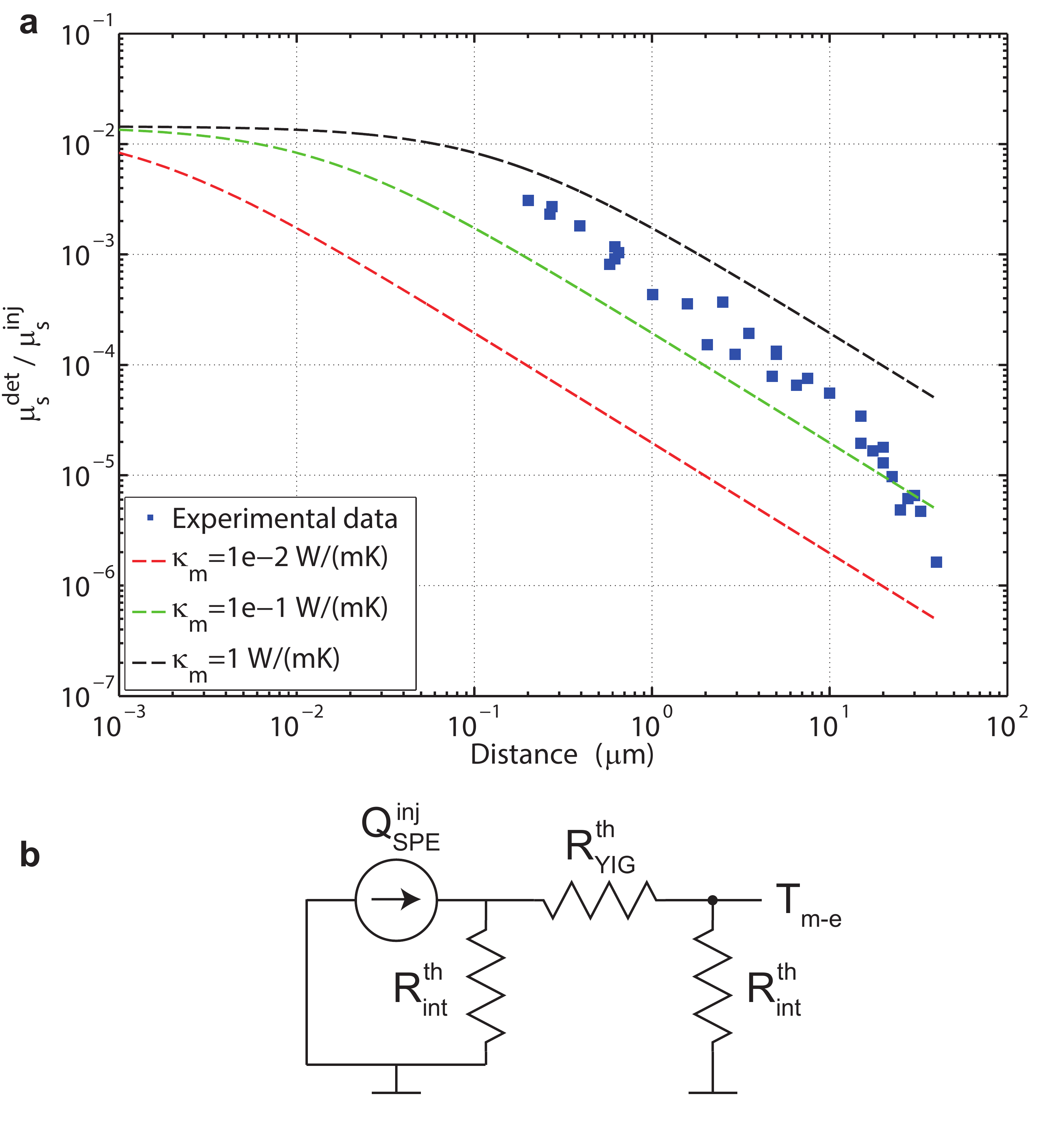}
\caption{(a) Results of the thermal model for $\kappa
_{m}=10^{-2}$ W/(mK) (red curve), $\kappa_{m}=10^{-1}$ W/(mK) (green curve)
and $\kappa_{m}=1$ W/(mK) (black curve). Plotted on the y-axis is the spin
transfer efficiency resulting from the thermal model, $\eta_{\mathrm{th}}%
=\mu_{s}^{\mathrm{det}}/\mu_{s}^{\mathrm{inj}}$. The blue squares represent
the experimental data. (b) The equivalent thermal resistance model. The
definitions of the thermal resistances used in the model are given in the main
text. At the thermal grounds in the circuit, the temperature difference
between magnons and electrons ($T_{m-e}$) is zero. }%
\label{fig:thermal_spin_transfer_efficiency}%
\end{figure}The equivalent circuit\ is based on the spin Peltier heat current
and the spin thermal resistances of the YIG%
$\vert$%
Pt interfaces and the YIG channel. This allows us to find $T_{\mathrm{m-e}}$,
the temperature difference between magnons and electrons at the detector
interface, which is the driving force for the SSE in this model. The
equivalent thermal resistance circuit is shown in
Fig.~\ref{fig:thermal_spin_transfer_efficiency}(b). Relaxation is disregarded,
so the model is only valid for $d<\ell_{\mathrm{mp}}$. The interface magnetic
heat resistance is given by $R_{\mathrm{int}}^{\mathrm{th}}=1/(\kappa_{s}%
^{I}A_{\mathrm{int}})$, with $\kappa_{s}^{I}$ equal to \cite{Xiao2010,
Flipse2014, PhysRevB.88.094410}
\begin{equation}
\kappa_{s}^{I}=\frac{h}{e^{2}}\frac{k_{B}T}{\hbar}\frac{\mu_{B}k_{B}%
g^{\uparrow\downarrow}}{\pi M_{s}\Lambda^{3}},
\end{equation}
and where $\mu_{B}$ is the Bohr magneton. The YIG heat resistance
$R_{\mathrm{YIG}}^{\mathrm{th}}=d/(\kappa_{m}A_{\mathrm{YIG}})$ and from the
thermal circuit model we find that $T_{\mathrm{m-e}}=Q_{\mathrm{SPE}%
}^{\mathrm{inj}}\left(  R_{\mathrm{int}}^{\mathrm{th}}\right)  ^{2}/\left(
R_{\mathrm{int}}^{\mathrm{th}}+R_{\mathrm{YIG}}^{\mathrm{th}}\right)  $, which
generates a spin accumulation in the detector by the spin Seebeck effect
\begin{equation}
\mu_{s}^{\mathrm{det}}=T_{\mathrm{m-e}}\frac{g^{\uparrow\downarrow}\gamma\hbar
k_{B}}{\pi M_{s}\Lambda^{3}}\frac{4\pi}{e}\frac{\ell_{s}}{\sigma}\tanh\left(
\frac{t}{2\ell_{s}}\right)  \frac{1+e^{-2t/\ell_{s}}}{\left(  1-e^{-t/\ell
_{s}}\right)  ^{2}}.
\end{equation}
The thus obtained spin transfer efficiency $\eta_{\mathrm{th}}$ is plotted in
Fig.~\ref{fig:thermal_spin_transfer_efficiency}(a) as a function of the magnon
spin conductivity $\kappa_{m}$. For $\kappa_{m}\sim0.1-1$ W/(mK) reasonable
agreement with the experimental data can be achieved. While Schreier \emph{et
al.} argued that $\kappa_{m}$ should be in the range $10^{-2}-10^{-3}$
W/(mK)), $\kappa_{m}$ from Tab.~\ref{tab:transportcoefficients} is also of the
order of 1 W/(mK) at room temperature. Hence, the magnon temperature model can
describe the non-local experiments, provided that the magnon-phonon relaxation
length $\ell_{\mathrm{mp}}$ is large. However, from the expression for
$\ell_{\mathrm{mp}}$ that we gave in Tab.~\ref{tab:transportcoefficients} we
find that $\ell_{\mathrm{mp}}\sim10$ $\mathrm{\mu}$m corresponds to
$\tau_{\mathrm{mp}}\approx\tau_{\mathrm{mr}}\sim1$ ns and $\kappa_{m}%
\sim10^{4}$ W/(mK), which is at least three orders of magnitude larger than
even the total YIG heat conductivity, and is clearly unrealistic. Thus,
requiring $\ell_{\mathrm{mp}}\sim10$ $\mathrm{\mu}$m while maintaining
$\kappa_{m}\sim1$ W/(mK) is inconsistent. Also, an $\ell_{\mathrm{mp}}$ of the
order of nanometers as reported by Schreier \emph{et al.} and Flipse \emph{et
al.} is difficult to reconcile with the observed length scale of the order of
10 $\mathrm{\mu}$\textrm{m. }

Up to now we disregarded phononic heat transport. As argued, the interaction
of phonons with magnons in the spin channel is weak, but the energy transfer
can be efficient. The spin Peltier effect at the contact generates a magnon
heat current that decays on the length scale $\ell_{\mathrm{mp}}$, heating up
the phonons that subsequently diffuse to the detector, where they cause a spin
Seebeck effect. The magnon system is in equilibrium except at distances from
injector and detector on the scale $\ell_{\mathrm{mp}}$ that we argued to be
short. In this scenario there is no non-local magnon transport in the bulk at all, but
injector and detector communicate by pure phonon heat transport. However, this
mechanism does not explain the exponential decay of the non-local signal: the
diffusive heat current emitted by a line source, taking into account that the GGG
substrate has a heat conductivity close to that of YIG
\cite{PhysRevB.88.094410}, decays only logarithmically as a function of distance.

\subsection{Longitudinal spin Seebeck effect}

\label{subsec:sse} The spin Seebeck effect is usually measured in the
longitudinal configuration, i.e. samples with a YIG film grown on gadolinium
gallium garnet (GGG) and a Pt top contact, for which our one-dimensional model
\cite{2015arXiv150501329D} applies.\ A recent study extracted the length
scale of the longitudinal spin Seebeck effect from experiments on samples with
various YIG film thicknesses \cite{Kehlberger2015}. A length of the order of 1
$\mathrm{\mu}$m was found. Similar results were obtained by Kikkawa \emph{et al.} \cite{Kikkawa2015}.

\begin{figure}[ptb]
\includegraphics[width=8.5cm]{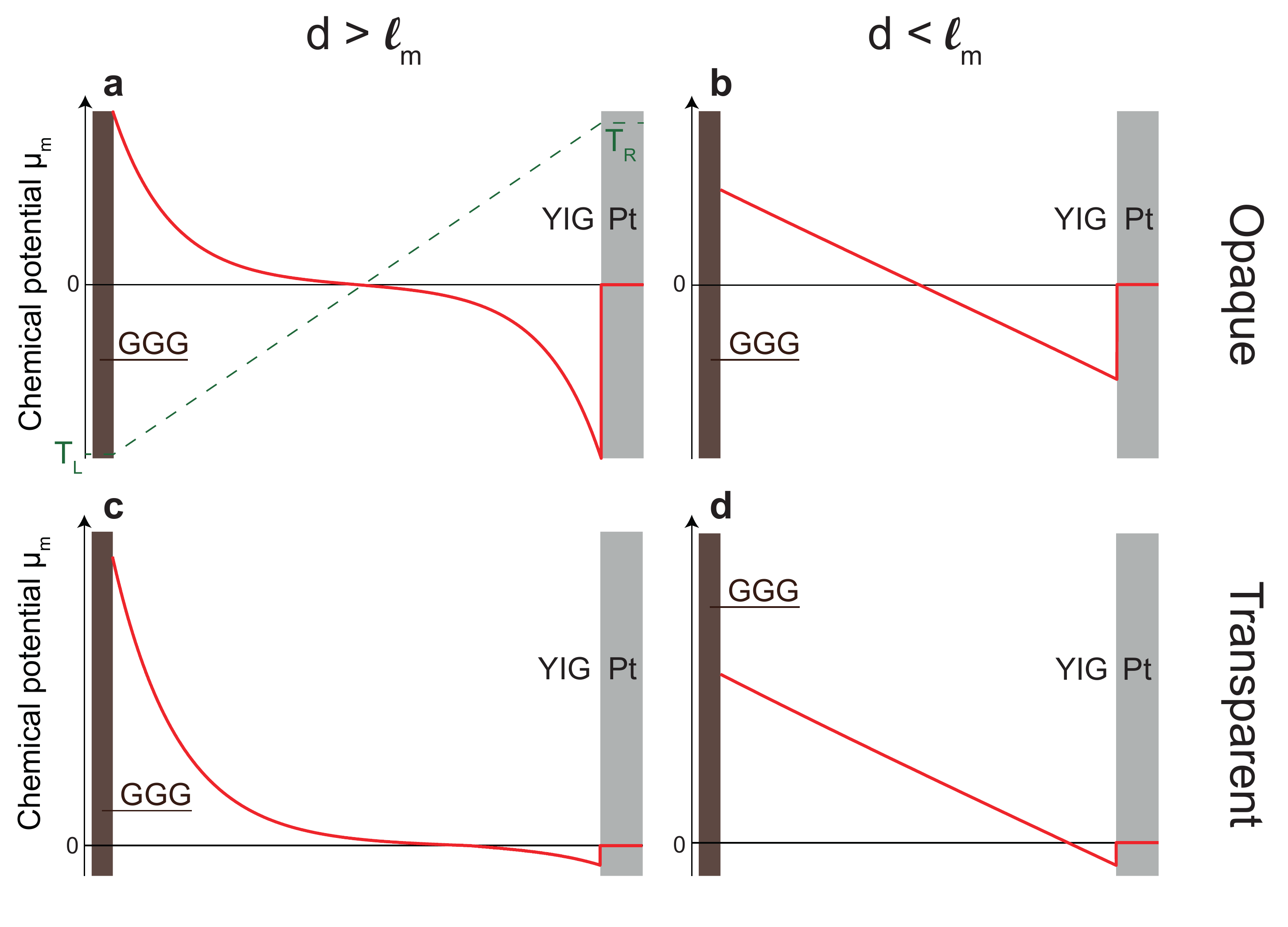} \caption{Magnon chemical potential $\mu_{m}$ under the spin Seebeck effect for a linear
temperature gradient in YIG, in the limit of: (a) an opaque interface and
thick YIG, (b) an opaque interface and thin YIG, (c) a transparent interface
and thick YIG and (d) a transparent interface and thin YIG. In all four cases,
$\mu_{m}$ changes sign somewhere in the YIG. For higher interface transparency
(larger $g_{s}$), the zero point shifts closer to the Pt$|${}YIG interface.}%
\label{fig:SSE_limits}%
\end{figure}

We assume a constant gradient $(T_{L}-T_{R})/d<0,$ where $T_{L},T_{R}$ are the
temperatures at the interfaces of YIG to GGG,platinum, respectively, with
$T_{m}\ $everywhere equilibrized to $T_{p},$ and disregard the Kapitza heat resistance, cf. Fig.~\ref{fig:SSE_limits}(a).
At the YIG%
$\vert$%
GGG interface the spin current vanishes. Figs.~\ref{fig:SSE_limits} illustrate
the magnon chemical potential profile on the YIG thickness $d$ as well as the
transparency of the Pt%
$\vert$%
YIG interface for four limiting cases, i.e. for opaque ($g_{s}<\sigma_{m}%
/\ell_{m}$) and transparent ($g_{s}>\sigma_{m}/\ell_{m}$) interfaces and a
thick ($d>\ell_{m}$) and a thin ($d<\ell_{m}$) YIG film, in which analytic
results can be derived.

We define a spin Seebeck coefficient as the normalized inverse spin Hall
voltage $V_{\mathrm{ISHE}}/t_{y}$ in the platinum film of length $t_{y}$
divided by the temperature gradient $\Delta T/d,$ with $\Delta T=T_{L}-T_{R}$
and average temperature $T_{0}$:
\begin{equation}
\sigma_{\mathrm{SSE}}=\frac{dV_{\mathrm{ISHE}}}{t_{y}\Delta T}.
\end{equation}
\begin{figure}[h]
\centering
\includegraphics[width=8.5cm]{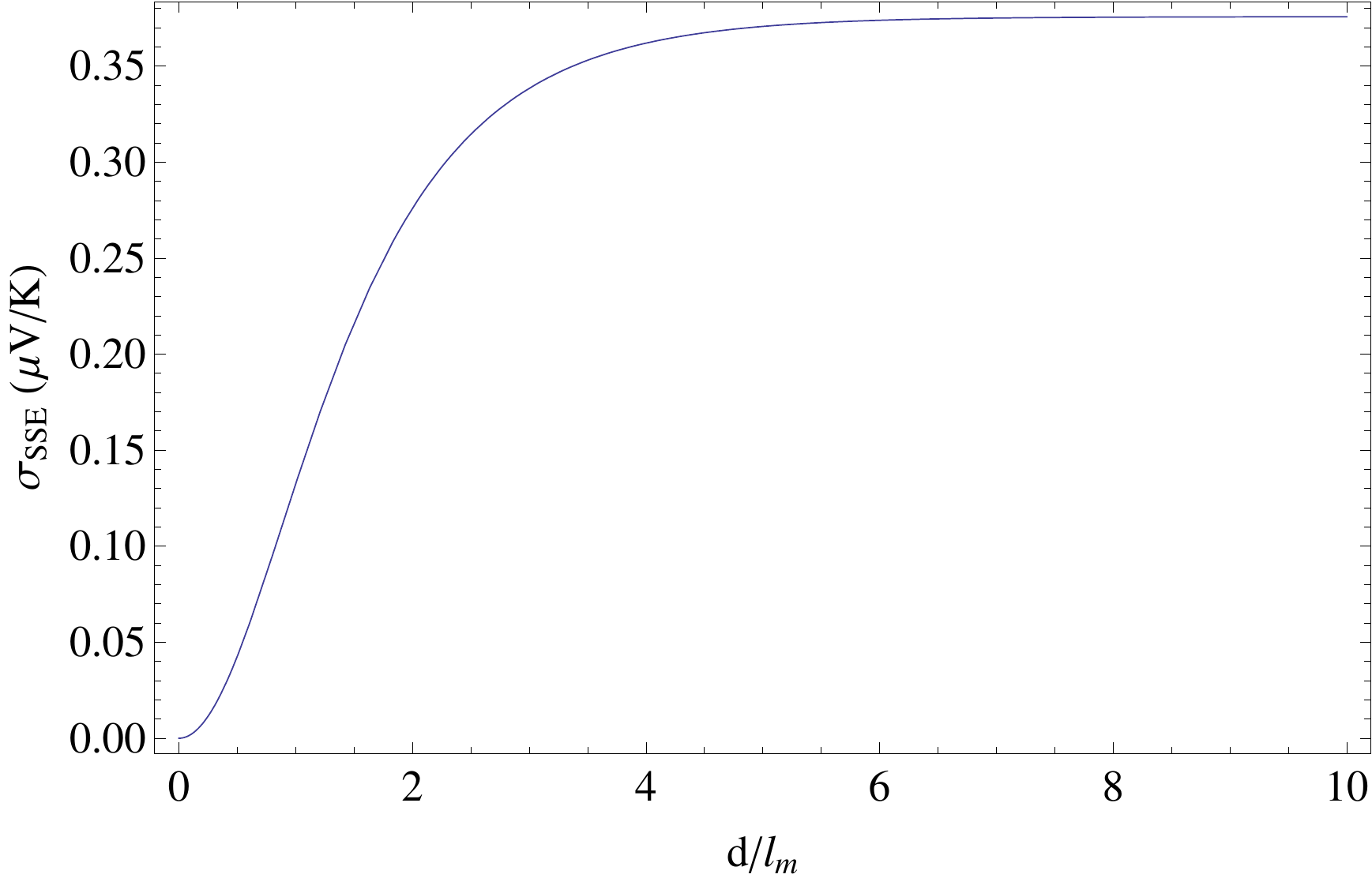} \caption{Normalized spin
Seebeck coefficient as a function of the thickness of the magnetic insulator
in the direction of the temperature gradient. Parameters taken are from
Tab.~\ref{tab:modelparameters}, together with a Pt thickness of $t=10$ nm and
temperature of $300$ K. The value for the bulk spin Seebeck coefficient $L$ is
taken from the expression in Tab.~\ref{tab:transportcoefficients} with
$\tau=0.1$ ps.}%
\label{fig:sigmasse}%
\end{figure}Assuming that the Pt spin diffusion length $\ell_{s}$ is much
shorter than its film thickness $t$ we find the analytic expression
\begin{equation}
\sigma_{\mathrm{SSE}}=\frac{g_{s}\ell_{s}\ell_{m}L\theta\left[  \cosh\frac
{d}{\ell_{m}}-1\right]  }{t\sigma_{e}T_{0}\left[  g_{s}\ell_{m}\cosh\frac
{d}{\ell_{m}}+\sigma_{m}\left(  1+\frac{2g_{s}\ell_{s}}{\sigma_{e}}\right)
\sinh\frac{d}{\ell_{m}}\right]  }~.
\end{equation}
In Fig.~\ref{fig:sigmasse} $\sigma_{\mathrm{SSE}}$ is plotted as a function of
the relative thickness $d/\ell_{m}$ of the magnetic insulator in the transport
direction, Pt thickness of $t=10$ nm and $T_{0}=300$ K. We adopt $L$ from
Table~\ref{tab:transportcoefficients} and a relaxation time $\tau\sim
\tau_{\mathrm{mp}}\sim0.1$ ps and the parameters from Tab.~\ref{fig:sigmasse}.
The normalized spin Seebeck coefficient saturates as a function of $d$ on the
scale of the magnon spin diffusion length $\ell_{m}$. While experiments at
$T_0\leq250\,$K report somewhat smaller length scales than our $\ell_{m},$ our
saturation $\sigma_{\mathrm{SSE}}\sim0.1-1$ $\mathrm{\mu}$V/K is of the same
order as the experiments \cite{2015arXiv150606037G}.

In the limit of an opaque interface, $\sigma_{\mathrm{SSE}}$ saturates to
\begin{equation}
\sigma_{\mathrm{SSE}}(d\gg\ell_{m})=\frac{g_{s}\ell_{s}\ell_{m}L\theta}%
{tT_{0}\sigma_{e}\sigma_{m}}=\left(  \frac{g_{s}\ell_{s}}{\sigma_{e}}\right)
\left(  \frac{\ell_{m}}{t}\right)  \frac{\alpha_{\mu}\theta k_{B}}{e}~,
\end{equation}
in terms of the dimensionless ratio $\alpha_{\mu}$ from
Eq.~(\ref{eq:fullsteadystatemagnontransport}).

For a transparent interface with $\ell_{m}\gg\ell_{s}$ and $\sigma_{m}%
\sim\sigma_{e}$, the result is governed by bulk parameters only:%
\begin{equation}
\sigma_{\mathrm{SSE}}(d\rightarrow\infty)=\frac{\ell_{s}L\theta}{tT_{0}%
\sigma_{e}}~.
\end{equation}

This model for the spin Seebeck effect is oversimplified by assuming a
vanishing magnon-phonon relaxation length and disregarding interface heat
resistances. The gradient in the phonon temperature can give rise to a spin
Seebeck voltage \cite{Cornelissen2016} even when bulk magnon spin transport is
frozen out by a large magnetic field. Nevertheless, it is remarkable that it
gives a reasonable qualitative description for the spin Seebeck effect with
input parameters adapted for electrically-driven magnon transport. We conclude
that also in the description of the spin Seebeck effect the magnon chemical
potential can play a crucial role.

\section{Conclusions}

\label{sec:conclusions} We presented a diffusion theory for magnon spin and
heat transport in magnetic insulators actuated by metallic contacts. In
contrast to previous models, we focus on the magnon chemical potential. This
is an essential ingredient because under ambient conditions $\ell_{m}%
>\ell_{\mathrm{mp}}$, i.e., the magnon chemical potential relaxes over much
larger length scales than the magnon temperature.
We compare theoretical results for electrical magnon injection and detection
with non-local transport experiments on YIG%
$\vert$%
Pt structures \cite{Cornelissen2015}, for both a 1D analytical and a 2D
finite-element model.

In the 1D model we study the relevance of interface- vs. bulk-limited
transport and find that, for the materials and conditions considered, the
interface spin resistance dominates. For the limiting cases of transparent and
opaque interfaces the spin transfer efficiency $\eta$ decays algebraically
$\propto1/d$ as a function of injector-detector distance $d$ when $d<\ell_{m}%
$, and exponentially with a characteristic length $\ell_{m}$ for $d>\ell_{m}$.


A 2D finite element model for the actual sample configurations can be fitted
well to the experiments for different contact distances, leading to a magnon
conductivity $\sigma_{m}=5\times10^{5}$~S/m and diffusion length $\ell
_{m}=9.4~\mathrm{\mu}$m.

The experiments measure first and second order harmonic signals that are
attributed to electrical magnon spin injection/detection and thermal
generation of magnons by Joule heating with spin Seebeck effect detection,
respectively. Here, we focus on the linear response that we argue to be
dominated by the diffusion of a magnon accumulation governed by the chemical
potential, rather than the magnon temperature. However, we applied our theory
also to the standard longitudinal (local) spin Seebeck geometry. We find the
same length scale $\ell_{m}$ and a (normalized) spin Seebeck coefficient of
$\sigma_{\mathrm{SSE}}\sim0.1-1$ $\mathrm{\mu}$V/K for $d\gg\ell_{m},$ which
is of the same order of magnitude as the observations \cite{Kehlberger2015}.

\acknowledgements We would like to acknowledge H. M. de Roosz and J.G.
Holstein for technical assistance, and Yaroslav Tserkovnyak, Arne Brataas,
Scott Bender, Jiang Xiao, and Benedetta Flebus for discussions. This work is
part of the research program of the Foundation for Fundamental Research on
Matter (FOM) and supported by NanoLab NL, EU FP7 ICT Grant No. 612759 InSpin,
Grant-in-Aid for Scientific Research (Grant Nos. 25247056, 25220910, 26103006)
and the Zernike Institute for Advanced Materials. RD is member of the D-ITP
consortium, a program of the Netherlands Organization for Scientific Research
(NWO) that is funded by the Dutch Ministry of Education, Culture and Science (OCW).

\appendix

\section{Boltzmann transport theory}

\label{sec:appendix} Here we derive our magnon transport theory from the
linearized Boltzmann equation in the relaxation time approximation, thereby
introducing and estimating the different collision times.

\subsection{Boltzmann equation}

Eqs.~(\ref{eq:linearresponsespinheatmagnet},\ref{eq:diffeqsmagnet}%
,\ref{eq:fullsteadystatemagnontransport}) are based on the Boltzmann equation
for the magnon distribution function $f(\mathbf{x},\mathbf{k},t)$:
\begin{equation}
\frac{\partial f}{\partial t}+\frac{\partial f}{\partial\mathbf{x}}\cdot
\frac{\partial\omega_{\mathbf{k}}}{\partial\mathbf{k}}=\Gamma^{\mathrm{in}%
}[f]-\Gamma^{\mathrm{out}}[f]~,\label{eq:boltzmannfull}%
\end{equation}
where $\Gamma^{\mathrm{in}}=\Gamma_{\mathrm{el}}^{\mathrm{in}}+\Gamma
_{\mathrm{mr}}^{\mathrm{in}}+\Gamma_{\mathrm{mp}}^{\mathrm{in}}+\Gamma
_{\mathrm{mm}}^{\mathrm{in}}$ and $\Gamma^{\mathrm{out}}=\Gamma_{\mathrm{el}%
}^{\mathrm{out}}+\Gamma_{\mathrm{mr}}^{\mathrm{out}}+\Gamma_{\mathrm{mp}%
}^{\mathrm{out}}+\Gamma_{\mathrm{mm}}^{\mathrm{out}}$ are the total rates of
scattering into and out of a magnon state with wave vector $\mathbf{k}$,
respectively. The subscripts refer to elastic magnon scattering at defects,
magnon relaxation by magnon-phonon interaction that do not conserve magnon
number, magnon-conserving inelastic and elastic magnon-phonon interactions,
and magnon number and energy-conserving magnon-magnon interactions. We discuss
them in the following for an isotropic magnetic insulator and in the limit of
small magnon and phonon numbers.

The elastic magnon scattering is given by Fermi's Golden rule as
\begin{equation}
\Gamma_{\mathrm{el}}^{\mathrm{out}}=\frac{2\pi}{\hbar}\sum_{\mathbf{k}%
^{\prime}}\left\vert V_{\mathbf{k}\mathbf{k}^{\prime}}^{\mathrm{el}%
}\right\vert ^{2}\delta(\hbar\omega_{\mathbf{k}}-\hbar\omega_{\mathbf{k}%
^{\prime}})f(\mathbf{k},t)~,
\end{equation}
where $V_{\mathbf{k}\mathbf{k}^{\prime}}^{\mathrm{el}}$ is the matrix element
for scattering by defects and rough boundaries
\cite{PhysRev.152.731,2015arXiv151005316F} of a magnon with momentum
$\hbar\mathbf{k}$ to one with $\hbar\mathbf{k}^{\prime}$ at the same energy.
$\Gamma_{\mathrm{el}}^{\mathrm{in}}$ is obtained from this expression by
interchanging $\mathbf{k}$ and $\mathbf{k}^{\prime}$. In the presence of the
in-scattering term (vertex correction) $\Gamma_{\mathrm{el}}^{\mathrm{in}}$
the Boltzmann equation is an integrodifferential rather than a simple
differential equation.

Gilbert damping parameterizes the magnon dissipation into the phonon bath.
According to the linearized Landau-Lifshitz-Gilbert equation
\cite{PhysRevB.90.094409}
\begin{equation}
\Gamma_{\mathrm{mr}}^{\mathrm{out}}=2\alpha_{G}\omega_{\mathbf{k}}%
f(\mathbf{k},t).
\end{equation}
Since the phonons assumed at thermal equilibrium with temperature $T_{p}$,
$\Gamma_{\mathrm{mr}}^{\mathrm{in}}$ is obtained by substituting
$f(\mathbf{k},t)\rightarrow n_{B}\left(  \hbar\omega_{\mathbf{k}}/k_{B}%
T_{p}\right)  $ in $\Gamma_{\mathrm{mr}}^{\mathrm{out}}$.

Magnon-conserving magnon-phonon interactions with matrix elements
$V_{\mathbf{k}\mathbf{k}^{\prime}\mathbf{q}}^{\mathrm{mp}}$ generate the
out-scattering rate
\begin{align}
\Gamma_{\mathrm{mp}}^{\mathrm{out}}  &  =\frac{2\pi}{\hbar}\sum_{\mathbf{k}%
^{\prime},\mathbf{q}}\left\vert V_{\mathbf{k}\mathbf{k}^{\prime}\mathbf{q}%
}^{\mathrm{mp}}\right\vert ^{2}\delta(\hbar\omega_{\mathbf{k}}-\hbar
\omega_{\mathbf{k}^{\prime}}-\epsilon_{\mathbf{q}}%
)\nonumber\label{eq:mpscatteringrate}\\
&  \times f(\mathbf{k},t)((1+f(\mathbf{k}^{\prime},t))\left[  1+n_{B}\left(
\frac{\epsilon_{\mathbf{q}}}{k_{B}T_{p}}\right)  \right]  ~,
\end{align}
where $\epsilon_{\mathbf{q}}=\hbar c|\mathbf{q}|$ is the acoustic phonon
dispersion with sound velocity $c$ and momentum $\mathbf{q.}$ The
\textquotedblleft in\textquotedblright\ scattering rate
\begin{align}
\Gamma_{\mathrm{mp}}^{\mathrm{in}}  &  =\frac{2\pi}{\hbar}\sum_{\mathbf{k}%
^{\prime},\mathbf{q}}\left\vert V_{\mathbf{k}\mathbf{k}^{\prime}\mathbf{q}%
}^{\mathrm{mp}}\right\vert ^{2}\delta(\hbar\omega_{\mathbf{k}}-\hbar
\omega_{\mathbf{k}^{\prime}}-\epsilon_{\mathbf{q}}%
)\nonumber\label{eq:mpinteractionrate}\\
&  \times f(\mathbf{k}^{\prime},t)((1+f(\mathbf{k},t))n_{B}\left(
\frac{\epsilon_{\mathbf{q}}}{k_{B}T_{p}}\right)  ~.
\end{align}
Finally, the four-magnon interactions (two magnons in, two magnons out)
generate
\begin{align}
\Gamma_{\mathrm{mm}}^{\mathrm{out}}  &  =\frac{2\pi}{\hbar}\sum_{\mathbf{k}%
^{\prime},\mathbf{k}^{\prime\prime},\mathbf{k}^{\prime\prime\prime}}\left\vert
V_{\mathbf{k}+\mathbf{k}^{\prime},\mathbf{k}-\mathbf{k}^{\prime}%
,\mathbf{k}^{\prime\prime}-\mathbf{k}^{\prime\prime\prime}}^{\mathrm{mm}%
}\right\vert ^{2}\nonumber\label{eq:mmscatteringrate}\\
&  \times\delta(\hbar\omega_{\mathbf{k}}+\hbar\omega_{\mathbf{k}^{\prime}%
}-\hbar\omega_{\mathbf{k}^{\prime\prime}}-\hbar\omega_{\mathbf{k}%
^{\prime\prime\prime}})\delta(\mathbf{k}+\mathbf{k}^{\prime}-\mathbf{k}%
^{\prime\prime}-\mathbf{k}^{\prime\prime\prime})\nonumber\\
&  \times f(\mathbf{k},t)f(\mathbf{k}^{\prime},t)[1+f(\mathbf{k}^{\prime
\prime},t)][1+f(\mathbf{k}^{\prime\prime\prime},t)]~,
\end{align}
while $\Gamma_{\mathrm{mm}}^{\mathrm{in}}$ follows by exchanging $\mathbf{k}$
$\mathbf{k}^{\prime\prime}$, and $\mathbf{k}^{\prime}$ and $\mathbf{k}%
^{\prime\prime\prime}$. Disregarding umklapp scattering, the magnon-magnon
interactions conserve linear and angular momentum. $V^{\mathrm{mm}}$ therefore
depends only on the center-of-mass momentum and the relative magnon momenta
before and after the collision, which implies that $\Gamma_{\mathrm{mm}}$ does
not affect transport directly (analogous to the role of electron-electron
interactions in electric conduction).

The collision rates govern the energy and momentum-dependent collision times
$\tau_{a}(k,\hbar\omega)$ (with $a\in\{\mathrm{el},\mathrm{mr},\mathrm{mp}%
,\mathrm{mm}\}$). These are defined from the \textquotedblleft
out\textquotedblright\ rates via
\begin{equation}
\frac{1}{\tau_{a}(k,\hbar\omega)}=\frac{\Gamma_{a}^{\mathrm{out}}%
}{f(\mathbf{k},t)},
\end{equation}
replacing $f\rightarrow n_{B}(\hbar\omega_{\mathbf{k}}/k_{B}T_{p})$ and
$\hbar\omega_{\mathbf{k}}$ with $\hbar\omega$ where phonons are involved. Here
we are interested mainly in thermal magnons for which the relevant collision
times are evaluated at energy $\hbar\omega=k_{B}T$ and momentum $k=\Lambda
^{-1}$. Then $1/\tau_{\mathrm{mr}}\sim\alpha_{G}k_{B}T/\hbar$. Elastic magnon
scattering can be parameterized by a mean-free-path $\ell_{\mathrm{el}}%
=\tau_{\mathrm{el}}(k,\hbar\omega)\partial\omega_{\mathbf{k}}/\partial k$, and
therefore $1/\tau_{\mathrm{el}}(k,\hbar\omega)=2\ell_{\mathrm{el}}^{-1}%
\sqrt{J_{s}\omega/\hbar}$ or $\tau_{\mathrm{el}}=\ell_{\mathrm{el}%
}/v_{\mathrm{m}}$, where $v_{\mathrm{m}}=2\sqrt{J_{s}\omega/\hbar}$ is the
magnon group velocity. Estimates for $\ell_{\mathrm{el}}$ range from 1
$\mathrm{\mu}$m \cite{2015arXiv151005316F} under the assumption that $\ell
_{m}$ is due to Gilbert damping and disorder only, to $500$ $\mathrm{\mu}$m
\cite{PhysRev.152.731}. Therefore $\tau_{\mathrm{el}}\sim10-10^{5}$ ps. Since
we deduce in the main text that at room temperature $\tau_{\mathrm{mp}}$ is
one to two orders of magnitude smaller than this $\tau_{\mathrm{el}}$, we
completely disregard elastic two-magnon scattering in the comparison with experiments.

We adopt the relaxation time approximation in which the scattering terms read
\begin{align}
\Gamma[f] &  =\frac{1}{\tau_{\mathrm{el}}}\left[
f-n_{B}\left(  \frac{\hbar\omega_{\mathbf{k}}-\mu_{m}}{k_{B}T_{m}}\right)
\right]  \nonumber\\
+ &  \frac{1}{\tau_{\mathrm{mr}}}\left[  f-n_{B}\left(  \frac{\hbar
\omega_{\mathbf{k}}}{k_{B}T_{p}}\right)  \right]  \nonumber\\
+ &  \frac{1}{\tau_{\mathrm{mp}}}\left[  f-n_{B}\left(  \frac{\hbar
\omega_{\mathbf{k}}-\mu_{m}}{k_{B}T_{p}}\right)  \right]  \nonumber\\
+ &  \frac{1}{\tau_{\mathrm{mm}}}\left[  f-n_{B}\left(  \frac{\hbar
\omega_{\mathbf{k}}-\mu_{m}}{k_{B}T_{m}}\right)  \right]
~.\label{eq:boltzmannreltimeapprox}%
\end{align}
The distribution functions here are chosen such that the elastic scattering
processes stop when $f$ approaches the Bose-Einstein distribution with local
chemical potential $\mu_{m}\neq0,$ in contrast to the inelastic scattering
that cause relaxation to thermal equilibrium with the lattice and $\mu_{m}=0.$
Similarly, the temperatures $T_{p}$ vs. $T_{m}$ are chosen to express that the
scattering exchanges energy with the phonons or keeps it in the magnon system, respectively.

The Boltzmann equation may be linearized in terms of the small perturbations,
i.e. the gradients of temperature and chemical potential. The local momentum
space shift $\delta f$ of the magnon distribution function
\begin{equation}
\delta f\left(  \mathbf{x},\mathbf{k}\right)  =\tau\frac{\partial n_{B}\left(
\frac{\hbar\omega_{\mathbf{k}}}{k_{B}T_{p}}\right)  }{\partial\hbar
\omega_{\mathbf{k}}}\frac{\partial\omega_{\mathbf{k}}}{\partial\mathbf{k}%
}\cdot\left(  \boldsymbol{\nabla}_{\mathbf{x}}\mu_{m}+\hbar\omega_{\mathbf{k}%
}\frac{\boldsymbol{\nabla}_{\mathbf{x}}T_{m}}{T_{p}}\right)  ~,
\label{eq:deltaf}%
\end{equation}
where $1/\tau=1/\tau_{\mathrm{mr}}+1/\tau_{\mathrm{mp}}$. The magnon spin and
heat currents Eq.~(\ref{eq:linearresponsespinheatmagnet}) are obtained by
substituting $\delta f$ into
\begin{align}
\mathbf{j}_{m}  &  =\hbar\int\frac{d\mathbf{k}}{\left(  2\pi\right)  ^{3}%
}\delta f\left(  \mathbf{k}\right)  \frac{\partial\omega_{\mathbf{k}}%
}{\partial\mathbf{k}}~,\label{eq:defspinandheatcurrents}\\
\mathbf{j}_{Q,m}  &  =\int\frac{d\mathbf{k}}{\left(  2\pi\right)  ^{3}}\delta
f\left(  \mathbf{k}\right)  \hbar\omega_{\mathbf{k}}\frac{\partial
\omega_{\mathbf{k}}}{\partial\mathbf{k}}~.
\end{align}

The magnon spin and heat diffusion Eqs.~(\ref{eq:diffeqsmagnet}) are obtained
by a momentum integral of the Boltzmann equation
(\ref{eq:boltzmannreltimeapprox}) after multiplying by $\hbar$ and
$\hbar\omega_{\mathbf{k}}$, respectively. The local distribution function in
the collision terms consists of the sum of the \textquotedblleft
drift\textquotedblright\ term $\delta f$ and the Bose-Einstein distribution
with local temperature and chemical potential
\begin{equation}
f(\mathbf{k},t)=\delta f+n_{B}((\hbar\omega_{\mathbf{k}}-\mu_{m}%
(\mathbf{x}))/k_{B}T_{m}(\mathbf{x})))
\end{equation}
We reiterate that the relatively efficient magnon conserving $\tau
_{m}$ limits the energy, but not (directly) the spin diffusion.

\subsection{Magnon-magnon scattering rate}

The four-magnon scattering rate is believed to efficiently thermalize the
local magnon distribution to the Bose-Einstein form
\cite{PhysRev.102.1217,PhysRevB.90.094409}. At room temperature the
leading-order correction to the exchange interaction in the presence of
magnetization textures reads
\begin{equation}
H_{\mathrm{xc}}=-\frac{J_{s}}{2s}\int d\mathbf{x}\mathbf{s}(\mathbf{x}%
)\cdot\nabla^{2}\mathbf{s}(\mathbf{x})~,
\end{equation}
where $\mathbf{s}(\mathbf{x})$ ($s=|\mathbf{s}|={S}/{a}^{3}%
$) is the spin density. By the Holstein-Primakoff transformation the spin
lowering operator reads $\hat{s}_{-}=s_{x}-is_{y}=\sqrt{2s-\hat{\psi}%
^{\dagger}\hat{\psi}}\hat{\psi}\simeq\sqrt{2s}\hat{\psi}-\hat{\psi}^{\dagger
}\hat{\psi}\hat{\psi}/2\sqrt{2s}$ in terms of the bosonic creation ($\hat
{\psi}^{\dagger}$) and annihilation ($\hat{\psi}$) operators. $H_{\mathrm{xc}%
}$ can be approximated as a four-particle point-like interaction term
\begin{equation}
H_{\mathrm{mm}}\approx g\int d\mathbf{x}\hat{\psi}^{\dagger}\hat{\psi
}^{\dagger}\hat{\psi}\hat{\psi}~,
\end{equation}
where $g\sim k_{B}T/s$ is the exchange interaction strength at thermal
energies. Using Fermi's Golden Rule for this interaction yields collision
terms as Eq.~(\ref{eq:mmscatteringrate}) with $V^{\mathrm{mm}}\approx g$:
\begin{align}
\frac{1}{\tau_{\mathrm{mm}}(k,\hbar\omega)} &  \approx\frac{g^{2}}{\hbar}%
\sum_{\mathbf{k}^{\prime},\mathbf{k}^{\prime\prime},\mathbf{k}^{\prime
\prime\prime}}\delta(\hbar\omega_{\mathbf{k}}+\hbar\omega_{\mathbf{k}^{\prime
}}-\hbar\omega_{\mathbf{k}^{\prime\prime}}-\hbar\omega_{\mathbf{k}%
^{\prime\prime\prime}})\nonumber\\
&  \times\delta(\mathbf{k}+\mathbf{k}^{\prime}-\mathbf{k}^{\prime\prime
}-\mathbf{k}^{\prime\prime\prime})\times n_{B}\left(  \frac{\hbar
\omega_{\mathbf{k}^{\prime}}}{k_{B}T_{p}}\right)  \nonumber\\
&  \left[  1+n_{B}\left(  \frac{\hbar\omega_{\mathbf{k}^{\prime\prime}}}%
{k_{B}T_{p}}\right)  \right]  \left[  1+n_{B}\left(  \frac{\hbar
\omega_{\mathbf{k}^{\prime\prime\prime}}}{k_{B}T_{p}}\right)  \right]  ~.
\end{align}
The momentum integrals can be estimated for thermal magnons with
$k=\Lambda^{-1}$ and $\hbar\omega=k_{B}T$ and
\begin{equation}
\frac{1}{\tau_{\mathrm{mm}}}\approx\frac{g^{2}}{\Lambda^{6}}\frac{{k_{B}T}%
}{\hbar}{\approx}\left(  \frac{T}{T_{c}}\right)  ^{3}\frac{{k_{B}T}}{\hbar}~,
\end{equation}
with Curie temperature $k_{B}T_{c}\approx J_{s}s^{2/3}$. With parameters for
YIG $J_{s}s^{2/3}/k_{B}\approx200$ K, which is the correct order of magnitude.
The $T^{4}$ scaling of the four-magnon interaction rate results from the
combined effects of the magnon density of states (magnon scattering phase
space) and energy-dependence of the exchange interactions.

While the magnon-magnon scattering is efficient at thermal energies, it
becomes slow at low energies close to the band edge due to phase space
restrictions and leads to deviations from the Bose-Einstein distribution
functions that may be disregarded at room temperature.

\subsection{Magnon-conserving magnon-phonon interactions}

At thermal energies and large wave numbers the magnon-conserving magnon-phonon
scattering \cite{PhysRev.152.731} is dominated by the dependence of the
exchange interaction on lattice distortions rather than magnetocrystalline
fields. Since we estimate orders of magnitude, we disregard phonon
polarization and the tensor character of the magnetoelastic interaction and
start from the Hamiltonian
\begin{equation}
H_{\mathrm{mp}}=-\frac{B}{s}\int d\mathbf{xs}(\mathbf{x})\cdot\nabla
^{2}\mathbf{s}(\mathbf{x})\left(  \sum_{\alpha\in\{x,y,z\}}\frac{\partial
R}{\partial x_{\alpha}}\right)  ~,
\end{equation}
where $B$ is a magnetoelastic constant. The scalar lattice displacement field
$R$ can be expressed in the phonon creation and annihilation operators
$\hat{\phi}^{\dagger}$ and $\hat{\phi}$ as
\begin{equation}
R=\sqrt{\frac{\hbar^{2}}{2\rho\epsilon}}\left[  \hat{\phi}+\hat{\phi}%
^{\dagger}\right]  ~,
\end{equation}
where $\epsilon$ is the phonon energy and $\rho$ the mass density. By the
Holstein-Primakoff transformation introduced in the previous section we find
to leading order
\begin{equation}
H_{\mathrm{mp}}\approx B\int dx\left(  \nabla\hat{\psi}^{\dagger}\right)
\cdot\left(  \nabla\hat{\psi}\right)  \left(  \frac{\hbar^{2}}{\rho\epsilon
}\right)  \left(  \sum_{\alpha\in\{x,y,z\}}\frac{\partial\hat{\phi}}{\partial
x_{\alpha}}\right)  +\mathrm{h.c.}%
\end{equation}
This Hamiltonian is the scattering potential in the matrix elements of
Eq.~(\ref{eq:mpinteractionrate})
\begin{equation}
\left\vert V_{\mathbf{k}\mathbf{k}^{\prime}\mathbf{q}}^{\mathrm{mp}%
}\right\vert ^{2}\approx\frac{B^{2}\hbar^{2}q^{2}}{\rho\epsilon_{\mathbf{q}}%
}\left(  \mathbf{k}\cdot\mathbf{k}^{\prime}\right)  ^{2}\delta(\mathbf{k}%
-\mathbf{k}^{\prime}-\mathbf{q})
\end{equation}
which by substitution and in the limit $\Lambda\ll\Lambda_{p}$, where
$\Lambda_{p}=\hbar c/k_{B}T_{p}$ is the phonon thermal de Broglie wavelength,
leads to
\begin{equation}
\frac{1}{\tau_{\mathrm{mp}}}\sim\frac{B^{2}}{\hbar\rho}\left(  \frac{\hbar
}{k_{B}T}\right)  ^{2}\frac{1}{\Lambda^{4}\Lambda_{p}^{5}}~,
\end{equation}
In the opposite limit $\Lambda\gg\Lambda_{p}$
\begin{equation}
\frac{1}{\tau_{\mathrm{mp}}}\sim\frac{B^{2}}{\hbar\rho}\left(  \frac{\hbar
}{k_{B}T}\right)  ^{2}\frac{1}{\Lambda^{7}\Lambda_{p}^{2}}~.
\end{equation}
At room temperature $\Lambda\approx\Lambda_{p}$ and for $\rho a^{3}=10^{-24}$
kg both expressions lead to $\tau_{\mathrm{mp}}=10(J_{s}/B)^{2}$ ns
\cite{PhysRevB.89.184413}. We could not find estimates of $B$ for YIG in the
literature. In iron, exchange interactions change by a factor of two upon
small lattice distortion $\Delta a\ll a$ \cite{PhysRevLett.83.2062}. While the
authors of this latter work find that this does not strongly affect the Curie
temperature, it leads to fast magnon-phonon scattering as we show now. Namely,
$B\sim a\left.  \partial J_{s}/\partial\Delta a\right\vert _{\Delta
a=0}\approx aJ_{s}/\Delta a$, so that $\tau_{\mathrm{mp}}=10(\Delta a/a)^{2}$
ns, which is many orders of magnitude smaller than one ns (and thus smaller
than $\tau_{\mathrm{mr}}$ at room temperature). While no proof, this argument
supports our hypothesis that the magnon temperature relaxation length is much
shorter than that of the magnon chemical potential.

\bibliography{main}

\end{document}